\documentclass[12pt]{article}
\pdfoutput=1
\usepackage{jheppub}
\def\be{\begin{eqnarray}}
\def\ee{\end{eqnarray}}

\usepackage{array}
\usepackage{amsmath}
\usepackage{multirow}
\usepackage{amssymb}
\usepackage{indentfirst}
\usepackage{pstricks}
\usepackage{titlesec}
\usepackage{tikz}
\usepackage{scalefnt}
\usepackage{pgfplots}
\numberwithin{equation}{section}
\titleformat{\section}
  {\normalfont\fontsize{16}{4}\bfseries}{\thesection}{1em}{}
  \usetikzlibrary{%
	calc,%
	decorations.pathmorphing,%
	fadings,%
	shadings%
}

\pgfplotsset{every axis/.append style={
font=\large,
line width=1pt,
tick style={line width=0.8pt}}}

\title{Conformal Bootstrap Approach to \\
$O(N)$ Fixed Points in Five Dimensions}

\author[]{Jin-Beom Bae}
\author[]{\ and Soo-Jong Rey}

\affiliation[]{School of Physics \& Center for Theoretical Physics\\
          Seoul National University, Seoul 151-747 \rm KOREA}

\abstract{
Whether $O(N)$-invariant conformal field theory exists in five dimensions with its implication to higher-spin holography was much debated. We find an affirmative result on this question by utilizing conformal bootstrap approach. In solving for the crossing symmetry condition, we propose a new approach based on specification for the low-lying spectrum distribution. We find the traditional one-gap bootstrapping is not suited since the nontrivial fixed point expected from large-$N$ expansion sits at deep interior (not at boundary or kink) of allowed solution region. We propose two-gap bootstrapping that specifies scaling dimension of two lowest scalar operators. The approach carves out vast region of lower scaling dimensions and universally features two tips. We find that the sought-for nontrivial fixed point now sits at one of the tips, while the Gaussian fixed point sits at the other tip. The scaling dimensions of scalar operators fit well with expectation based on large-$N$ expansion. We also find indication that the fixed point persist for lower values of $N$ all the way down to $N=1$. This suggests that interacting unitary conformal field theory exists in five dimensions for all nonzero $N$.
}


\preprint{}

\begin{document}
\maketitle

\section{Introduction}\label{intro}

The conformal bootstrap program \cite{Polyakov:1974gs} is a nonperturbative approach for solving conformal field theories, whose remarkable success is hallmarked by the complete solution of rational conformal field theories in two dimensional spacetime \cite{Belavin:1984vu}.
Recently, this program has been extended to higher dimensions with impressive results, especially, in its new approach to strongly interacting sectors \cite{Rattazzi:2008pe}- \cite{Rattazzi:2010gj}
and to solving quantum field theories with the best available computational efficiency and numerical precisions \cite{ElShowk:2012ht}, \cite{El-Showk:2014dwa}.
So far, all of these studies were confined to spacetime dimensions four or less.

The purpose of this paper is to extend the conformal bootstrap approach to conformal field theories in spacetime dimensions higher than four and explicitly demonstrate its promising utility by identifying interacting conformal field theories in five dimensions.

It has long been suspected that there are interacting conformal field theories in dimensions higher than 4. Yet, their existence were not clearly identified and even worse no theoretically satisfactory approach for systematic study of them was not developed. Recently, sparked by the suggestion from higher-dimensional higher-spin holography \cite{Joung:2012a} that extends the previous proposal of higher-spin holography in four dimensions
\cite{Sezgin:2002rt}, \cite{Klebanov:2002a}
and three dimensions \cite{Henneaux:2010xg}, there has been renewed interest to this question for theories that admit large $N$ expansion. In particular, for theories with $O(N)$ global symmetry, perturbative approach based on combined $1/N$- and $\epsilon$-expansions  \cite{Klebanov:2014a}, \cite{Klebanov:2014b} found positive indication for nontrivial ultraviolet fixed points. The result is very interesting and calls us for a better approach to overcome the perturbative nature of the method used. The conformal bootstrap program is one such approach and already aspects regarding central charge was studied \cite{Nakayama:2014a}.

The bootstrap approach utilizes internal consistency conditions that follow from rigid symmetries of (super)Poincar\'e, (super)conformal, and internal types, unitarity and crossing invariance. The simplest nontrivial boostrap condition is provided by the four-point correlation functions and were studied extensively starting from the early developments. To demonstrate utility of the conformal bootstrap program in higher dimensions, we explicitly work out for $O(N)$ symmetric four-point correlation functions in five dimensions.

The study required us to refine the approaches that have been widely used in four or less dimensions. Namely, in solving the crossing symmetry conditions, traditional approach specified the theory with the scaling dimension of the lowest non-identity operator $\Delta_{\rm min}$ (above the unitarity bound). We found that this traditional approach is incapable of locating nontrivial conformal field theory since, even for large $N$, the nontrivial ultraviolet fixed point predicted by $1/N$-expansion was swamped inside the allowed region in the space of low-lying scaling dimensions. This is in sharp contrast to the results in dimensions for our less, where interesting nontrivial fixed points (as well as the Gaussian fixed point) was located at the boundary of the allowed solution region.

In this traditional approach, no further specification of the spectrum is given: spectrum of the scaling dimensions would form some dense distribution above $\Delta_{\rm min}$. We borrow the terminology of band theory in solid-state physics and name $\Delta_{\rm min}$ as the band-gap \footnote{If the band-gap is of the same order as the level spacing in the spectrum above the band-gap, it is metallic. If the band-gap is much larger than the level spacing, it is insulating.}.
To remedy the situation and to be able to locate nontrivial fixed points at the boundary of allowed solution region, we propose to specify the theory by two lowest scaling dimensions $(\Delta_{\rm min}, \Delta_{\rm gap})$ of non-identity scalar operators. The lowest scaling dimension is at $\Delta_{\rm min}$ (above the unitarity bound) and the second lowest scaling dimension is further gapped at $\Delta_{\rm gap} > \Delta_{\rm min}$. Again, borrowing the terminology of band theory, we refer $\Delta_{\rm gap}$ as the band-gap and $\Delta_{\rm min}$ as the mid-gap \footnote{The mid-gap is provided by doped impurities, and turn the insulating or conducting phases to semiconducting or semi-metallic phases.}.

Applying the proposed two-gap approach, we found that the solutions to the bootstrap condition formed a region more constrained that those that were found by the traditional one-gap approach. The boundary of allowed solution region took the shape of $\Sigma$ with two pronounced tip points. For all specifications of $(N, \Delta_{\rm gap})$ to the conformal field theory, We found that one tip is the Gaussian fixed point,  while the other tip is the nontrivial fixed point whose location matches exactly with the location predicted by the $1/N$-expansion result of \cite{Klebanov:2014a}, \cite{Klebanov:2014b}. Within the numerical precision, we found an indication that the nontrivial fixed point persists to all values of $N$, all the way down to $N=1$.

This paper is organized as follows. In section 2, we start with recapitulation of fundamentals of the conformal bootstrap approach. After recalling the one-gap approach and its incapability to locate nontrivial fixed point, we put forward our two-gap approach for solving the boostrap conditions. In section 3, we present our results and analysis. We gather various nontrivial features that are associated with our two-gap approach. In section 4, we summarize our results and discuss various issue for future study.
In appendix A, we present moduli space of $n$-point correlation functions in a conformally invariant system in arbitrary spacetime dimension.

%%%%%%%%%%%%%%%%%%%%%%%%%%%%%%%%%%%%%%%%%%%%%%%%%%%%%%%%%
\section{Conformal Bootstrap Approach}
In this section, we recall aspects of the conformal bootstrap approach relevant for foregoing discussions.

%--------------------------------------------------------
\subsection{Conformal Correlation Functions}
Consider a conformal field theory in $d$-dimensional spacetime. The generators of the $SO(d+1,1)$ Euclidean conformal algebra are Poincar\'e translation $P_\mu$, rotation $M_{\mu \nu}$, dilatation $D$, and special conformal translation $K^\mu$. The correlation functions measure response of the system as a function of separations to perturbations sourced by local operators, so they should transform covariantly under the $SO(d+1,1)$. The conformal algebra fixes the structure of 2-point and 3-point correlation functions completely. In turn, conformal field theories are completely specified by 2- and 3-point correlation functions.

Denote local operators as ${\cal O}_I$, where $I$ refers collectively to all quantum numbers of the operator. Choosing the basis of local operators in orthonormal basis so that the 2-point correlation functions read
\be
\left\langle \mathcal{O}_I(x) \mathcal{O}_J(y) \right\rangle = \frac{\delta_{IJ} }{|x-y|^{2\Delta_I}},
\ee
where $\Delta_I$ refer to the conformal scaling dimension of $I$-th operator, the 3-point correlation functions
\be
\left\langle \mathcal{O}_I (x) \mathcal{O}_J (y) \mathcal{O}_K (z) \right\rangle =
{ C_{IJK} \over \Large{
|x-y|^{\Delta_I + \Delta_J - \Delta_K} |y-z|^{\Delta_J + \Delta_K - \Delta_I} |z-x|^{\Delta_K + \Delta_I - \Delta_J}}}
\ee
are completely specified by the structure constants $C_{IJK}$. Owing to the conformal invariance, total set of these structure constants are encoded by the operator product expansions (OPE). The OPE is most compactly expressible in radial quantization by ordering two operators at two different radii (equivalently, conformal time). For instance, the OPE of two identical scalar operators ${\cal O}$ reads \cite{Ferrara:1971a}
\begin{equation}
\mathcal{O}(x) \times \mathcal{O}(0) \sim \sum_{\Delta,\ell} C_{\Delta,\ell} \Phi_{\Delta,\ell}(x),
\end{equation}
where the structure constants $C_{\Delta,\ell}$ are partial wave expansion coefficients and $\Phi_{\Delta,\ell}$ is the partial wave amplitudes. The partial wave amplitudes includes the set of conformal primary operators. The conformal invariance dictates that all multipole moments of the OPE are primary states and their conformal descendants. In conformal field theory, every operator product is organized by conformal primary operators and their descendants, which are labeled by conformal dimension $\Delta$ and spin $\ell$.

The 4- or higher-point correlation functions are not fully fixed by the conformal invariance. For instance, 4-point correlation function of local operators inserted at $x_1, x_2, x_3, x_4$ comes with two arbitrary degrees of freedom: the conformal cross-ratios (anharmonic ratios)of operator insertion points defined by
\begin{equation}
u := \frac{x_{12}^2 x_{34}^2}{x_{13}^2 x_{24}^2} \qquad \mbox{and} \qquad v := \frac{x_{14}^2 x_{23}^2}{x_{13}^2 x_{24}^2},
\label{cross_ratio}
\end{equation}
where $x_{ij} \equiv x_i-x_j$. For instance, for local operators of the same kind, ${\cal O}$, the 4-point correlation function takes the form
\begin{equation}
\left\langle \mathcal{O}(x_1) \mathcal{O}(x_2) \mathcal{O}(x_3) \mathcal{O}(x_4) \right\rangle = \frac{1}{(x_{12})^{2\Delta} (x_{34})^{2\Delta}} F(u,v),
\label{4point}
\end{equation}
Here, $F(u, v)$ is a scalar function .

By construction, $u, v$ are invariant under the conformal transformation. As such, the function $F(u, v)$ has vanishing conformal weight, so further inputs are needed in order to constrain it. The conformal invariance allows to evaluate the multi-point correlation function by a sequence of operator product expansion (OPE). For the 4-point correlation function ($\ref{4point}$), this is reduced effectively to the OPE of two partial wave operators $\Phi_{\Delta,\ell}$. This OPE gives rise to the  dependence on the square of the structure constant $C_{\Delta,\ell}$ and to a nontrivial function that depends on the conformal cross-ratios $u, v$. Therefore, the conformal partial wave expansion of the scalar function $F(u, v)$ takes the form
\begin{equation}
F(u,v) = \sum_{\Delta,\ell} C_{\Delta,\ell}^2 G_{\Delta,\ell}(u,v).
\label{FtoG}
\end{equation}
The function $G_{\Delta, \ell}(u, v)$ is referred as the conformal block. If the theory is unitary, the reflection positivity asserts that the partial-wave coefficient $C_{\Delta,\ell}$ is real and hence $C^2_{\Delta, \ell}$ is positive definite.

We can get more information about the conformal block $G_{\Delta, \ell}(u, v)$ from the underlying conformal symmetry, and is derivable from quadratic Casimir of the $SO(d,2)$ conformal algebra \cite{Dolan:2011a}. If the spacetime dimension is even, the conformal block has a closed form expression in terms of hypergeometric functions\cite{Dolan:2001a,Dolan:2011a}. If the spacetime dimension is odd, it is not known yet whether the conformal block is expressible in closed form. In numerical bootstrap approach, we do not actually need to have such closed form expressions,  since we can evaluate the conformal block from its recursion relations \cite{Rychkov:2012a}.

%-------------------------------------------------------
\subsection{Moduli Space of $n$-Points in Various Dimensions}
Before proceeding further, we would like to generalize consideration of the previous subsection and count the number of conformal cross-ratios (anharmonic ratios) that can appear in $n$-point correlation functions. Since we could not find the discussions in the literature and since it has potential applications in other contexts such as scattering amplitudes and null polygon Wilson loops, we include the result in this section. The proof is relegated to the appendix.

Mathematically, counting the number of conformal cross-ratios (anharmonic ratios) amounts to counting the dimension of moduli space of $n$-points in $d$-dimensional space modulo conformal transformations. The conformal group $SO(d+1,1)$ in $d$-dimension has the dimension $\frac{(d+2)(d+1)}{2}$. So, one might naively suppose that the dimension of this moduli space is
\be
{\rm dim} {\cal M}^{\rm conf}(n, \mathbb{R}^d) = n d-\frac{(d+2)(d+1)}{2}.
\ee
However, this is not quite correct except for sufficiently large $n$ for a given $d$. We tabulate the correct dimensions of the moduli space in Table 1. Fortuitously, for four-point correlation functions, $n=4$, the number of conformal cross-ratios is always 2 so long as the spacetime dimension is greater than $1$.

\setlength{\tabcolsep}{3pt}
\setlength{\extrarowheight}{1.5pt}
{
\begin{table}[ht]
\centering
\begin{tabular}{|c||c|c|c|c|c||c|}
 \hline
$d$ &  2 pt & 3 pt   & 4 pt  & 5 pt & 6 pt      & n($>6$) pt \\ \hline \hline

       & $P^\mu : 1$  & $D : 1$           &          &          &           &        \\
 1 & $K^\mu : 1$  & $M^{\mu \nu} : 0$ & $n-3=1$  & $n-3=2$  & $n-3=3$   & $n-3$  \\
       & $n-2 = 0$    & $n-3 = 0$         &          &          &           &        \\ \hline

       & $P^\mu : 2$  & $D : 1$           &          &          &           &        \\
 2 & $K^\mu : 2$  & $M^{\mu \nu} : 1$ & $2n-6=2$ & $2n-6=4$ & $2n-6=6$  & $2n-6$  \\
       & $2n-4 = 0$   & $2n-6 = 0$  &      &         &      &   \\ \hline

       & $P^\mu : 3$  & $D : 1$           &                    &           &            &  \\
 3 & $K^\mu : 3$  & $M^{\mu \nu} : 2$ & $M^{\mu \nu} : 1$  & $3n-10=5$ & $3n-10=8$  & $3n-10$  \\
       & $3n-6 = 0$   & $3n-9 = 0$        & $3n-10=2$          &           &            &   \\ \hline

       & $P^\mu : 4$  & $D : 1$           &                   &                   &            &  \\
 4 & $K^\mu : 4$  & $M^{\mu \nu} : 3$ & $M^{\mu \nu} : 2$ & $M^{\mu \nu} : 1$ & $4n-15=9$  & $4n-15$  \\
       & $4n-8 = 0$   & $4n-12 = 0$       & $4n-14 = 2$       & $4n-15 = 5$       &            &   \\ \hline

       & $P^\mu : 5$  & $D : 1$           &                   &                   &                    &  \\
 5 & $K^\mu : 5$  & $M^{\mu \nu} : 4$ & $M^{\mu \nu} : 3$ & $M^{\mu \nu} : 2$ & $M^{\mu \nu} : 1$  & $5n-21$  \\
       & $5n-10 = 0$  & $5n-15 = 0$       & $5n-18 = 2$       & $5n-20 = 5$       & $5n-21 = 9$        &   \\ \hline \hline
       & $P^\mu : d$  & $D : 1$             &                       &                     &                      &  \\
 d & $K^\mu : d$  & $M^{\mu \nu} : d-1$ & $M^{\mu \nu} : d-2$   & $M^{\mu \nu} : d-3$ & $M^{\mu \nu} : d-4$  & $n d-$  \\
       & $dn-2d $  & $dn-3d $         & $dn-4d+2$         & $dn-5d+5$         & $dn-6d+9$        &   $ \frac{(d+2)(d+1)}{2} $ \\
& $= 0$  & $= 0$         & $= 2$         & $=5$         & $= 9$        &  \\
\hline
\end{tabular}
\caption{\sl Dimension of moduli space for various cases. For sufficient large $n$, we have compact expression $n d - \frac{(d+2)(d+1)}{2}$ because of generators are fully used to fix the points. The 4-point correlation function has 2 degree of freedom when spacetime dimension larger than 2. Therefore, crossing symmetry constraint available even for the five-dimensional bootstrap.}
\label{DOF}
\end{table}
}
%

%--------------------------------------------------------
\subsection{Radial Representation of Conformal Block}
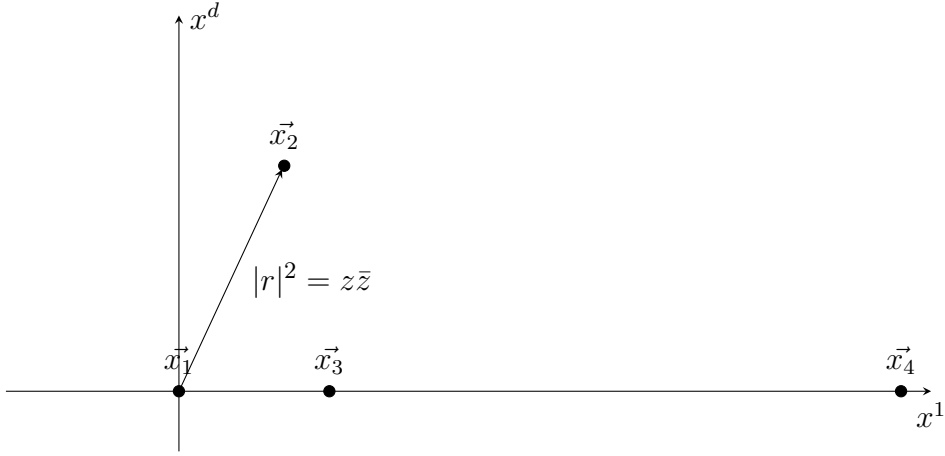
\begin{figure}[!]
  \centering
  \begin{tikzpicture}
   [
    scale=1,
    >=stealth,
    point/.style = {draw, circle,  fill = black, inner sep = 1.5pt},
    dot/.style   = {draw, circle,  fill = black, inner sep = .5pt},
  ]
     \coordinate (O) at (0,0);
     \draw[->] (-2.3,0) -- (10,0) coordinate[label = {below:$x^1$}] (xmax);
     \draw[->] (0,-0.8) -- (0,5) coordinate[label = {right:$x^d$}] (ymax);
     \draw[->] (0,0) -- (1.37,2.95) node[pos=0.5,right] {
  \fontsize{13pt}{0pt} {$|r|^2 = z \bar{z}$}};
    \node (n1) at (0,0) [point, label = above:$\vec{x_1}$] {};
    \node (n2) at (1.4,3) [point, label = above:$\vec{x_2}$] {};
    \node (n3) at (2,0) [point, label = above:$\vec{x_3}$] {};
    \node (n4) at (9.6,0) [point, label = above:$\vec{x_4}$] {};
  \end{tikzpicture}
  \caption{\sl Position of four insertion points of the sclar operators in (\ref{setting}). Using conformal symmetry, we fix $\vec{x_1}, \vec{x_3}, \vec{x_4}$. This leaves two degrees of freedom for the insertion point $\vec{x_2}$ lying in the $(x^1, x^D)$-subspace. The radial distance of $\vec{x_2}$ from origin is parametrized by $z,\bar{z}$. Therefore correlation function or conformal block is function of $z,\bar{z}$.}
  \label{fig_4point_setup}
\end{figure}

Denote the four points the local operators are inserted as $\vec{x_1}, \vec{x_2}, \vec{x_3}, \vec{x_4}$. Utilizing the conformal invariance, we can fix location of three points $\vec{x_1},\vec{x_3},\vec{x_4}$ as in Figure \ref{fig_4point_setup}. According to the result of previous subsection, there ought to be 2 remaining degree of freedom in arbitrary dimensions. Fixing 3 points as in figure \ref{fig_4point_setup} is consistent with this. Specifically, in five dimensions, we may conveniently put
\begin{equation}
\vec{x_1} = (0,0,0,0,0), \quad \vec{x_2} = (x_2^1,0,0,0,x_2^5), \quad \vec{x_3} = (1,0,0,0,0), \quad \vec{x_4} = \infty
\label{setting}
\end{equation}
Length $|\vec{x_{12}}|^2$ is $(x_2^1)^2+(x_2^5)^2$ for Euclidean space and $-(x_2^1)^2+(x_2^5)^2$ for Minkowski space. Therefore, we introduce two variables so that this length equals to $z\overline{z}$. In Euclidean space, the new variables $(z, \overline{z})$ are two complex variables related each other by complex conjugation. In Minkowski space, the new variables $(z, \overline{z})$ are two real-valued light-cone variables.

In terms of the new variables $(z, \overline{z})$, the cross-ratios (\ref{cross_ratio}) are given by
\be
u = z \overline{z} \qquad \mbox{and} \qquad v = (1-z)(1-\overline{z}).
\label{uv}
\ee
Being now a function of complex variables, the conformal block is in general a multi-valued function over the $z$-plane. It can be seen from the closed-form expressions of the conformal block in $d=2, 4$:
\begin{align}
G^{(d=4)}_{\Delta,\ell}(z, \overline{z}) &= \frac{(-1)^\ell}{2^\ell}
\frac{z\overline{z}}{z-\overline{z}}
\big[k_{\Delta + \ell}(z) k_{\Delta-\ell-2}(\overline{z}) - k_{\Delta + \ell}(\overline{z}) k_{\Delta-\ell-2}(z) \big] \nonumber \\
G^{(d=2)}_{\Delta,\ell}(z,\overline{z}) &= \frac{(-1)^\ell}{2^\ell}
\big[ k_{\Delta + \ell}(z) k_{\Delta-\ell}(\overline{z}) - k_{\Delta + \ell}(\overline{z}) k_{\Delta-\ell}(z) \big] \end{align}
where $k_\lambda (y)$ is the rescaled hypergeometric function:
\be
k_\lambda(z) & \equiv z^{\frac{\lambda}{2}}\ {_2F_1}\big[\frac{\lambda}{2},\frac{\lambda}{2},\lambda;z \big] .
\ee
We recall that the hypergeometric function has a cut $z \in [1, \infty)$ along the real axis.

To avoid aforementioned branch cuts and render the conformal block single-valued, we need to restrict $z, \overline{z}$ outside the cut along $z \in [1, \infty)$. This can be achieved by changing the variables $(z, \overline{z})$ to
\begin{equation}
\rho = \frac{z}{(1+\sqrt{1-z})^2} \qquad \mbox{and} \qquad \overline{\rho} = \frac{\overline{z}}{(1+\sqrt{1-\overline{z}})^2}.
\label{rad_trans}
\end{equation}
We are working in Euclidean space, so $(\rho, \overline{\rho})$ are complex conjugate each other.
Under the change of variable (\ref{rad_trans}), the $z$-plane outside the branch cut along $x \in [1, \infty)$ is mapped to the region inside a unit circle.

For the region inside a unit circle, we further change the variables to radial and polar variables:
\be
r = |\rho| \quad \mbox{and} \quad \eta = \cos (\mbox{arg}(\rho)), \qquad (0 \le r \le 1, \quad -1 \le \eta \le + 1).
\ee
The conformal block is now a function of $(r, \eta)$ in this bounded domain, so it can be expanded in double power series.
This expansion turns out to converge sufficiently fast \cite{Rychkov:2013a} and thus serve a useful basis for semi-definite programming.
The power series takes the form
\begin{equation}
G_{\Delta,\ell}(r,\eta) = \sum_{n=0}^{\infty} \sum_{j \in D(\ell)} B_{n,j}(\ell) r^{\Delta+n} {\Gamma(2\nu) \Gamma(j+1) \over \Gamma(2\nu + j)} %\frac{j!}{(2\nu)_j}
C_j^\nu(\eta)
\end{equation}
where $\nu = \frac{d}{2}-1$, $C_j^\nu(\eta)$ is the Gegenbauer polynomials, and the summation domain $D$ is given by
\be
D(\ell): \qquad j = \left\{ \begin{matrix}
0, 2 , 4, \cdots, \ell +n & \hskip1cm (\ell + n = 2 \mathbb{Z}) \\
1, 3, 5, \cdots, \ell + n & \hskip1cm (\ell + n = 2 \mathbb{Z} + 1)
\end{matrix} \right. .
\ee
The series coefficient $B_{n,j}$ is determined by the differential equation for the Casimir operator of the conformal algebra. It turns out the first component in radial expansion is given by
\begin{equation}
B_{0,j}(\ell) = 4^\Delta \delta_{j\ell}.
\end{equation}
Therefore, at leading order in radial expansion, the conformal block reads
\begin{equation}
G_{\Delta,\ell}(r,\eta) = (4r)^\Delta
{\Gamma(2\nu) \Gamma(\ell + 1) \over \Gamma(2\nu + \ell)}
C_\ell^\nu(\eta) + \mathcal{O}(r^{\Delta + 1})
\label{input}
\end{equation}

We take the crossing symmetric point $z=\bar{z}=\frac{1}{2}$, which corresponds to $r=3-2\sqrt{2}$. Higher order coefficients could be obtained similarly, but we do not need that information here. For more details of the coefficient $B_{n,j}$, we refer to \cite{Rychkov:2013a}.

One can compute the conformal block more efficiently by utilizing the Zamolodchikov recursive relation, as suggested in \cite{Simmons-Duffin:2013a}. It is reduced to a set of recursive relation given by
\begin{align}
h_{\Delta,\ell}(r,\eta) &\equiv r^{-\Delta} G_{\Delta, \ell}(r,\eta) \nonumber \\
h_{\Delta, \ell}(r,\eta) &= h_\ell^{\infty}(r,\eta) + \sum_i \frac{c_i r^{n_i}}{\Delta-\Delta_i} h_{\Delta_i + n_i,\ell_i}(r,\eta)
\label{radial_app}
\end{align}
Here, the term $h_\ell^{\infty}(r,\eta)$ refers to a  holomorphic function that specifies the `boundary condition' at $\Delta \rightarrow \infty$. This term can be  determined from the Sturm-Liouville problem of the quadratic Casimir operator of the conformal group and equals to
\begin{equation}
h_\ell^{\infty}(r,\eta) = \frac{\ell!}{(2\nu)_\ell} \frac{C_\ell^\nu(\eta)}{(1-r^2)^\nu \sqrt{(1+r^2)^2-4r^2 \eta^2}}.
\end{equation}
Detailed information of poles and coefficients $c_i$ can be found in the original work \cite{Simmons-Duffin:2013a}. In the foregoing analysis, we shall adopt this recursive derivation of the conformal block, taking (\ref{input}) as an input data. Numerical precision becomes better as the number of iteration is increased. In our run, we kept the precision up to $\sim 10^{-30}$ of exact values.

%--------------------------------------------------------
\subsection{One-Gap Bootstrapping: Review}

We are now at the stage of imposing the crossing symmetry and the unitarity. The conformal 4-point correlation function of same scalar operators is invariant under permutation of operator insertion points $x_1, x_2, x_3, x_4$. A nontrivial constraint follows from exchange of two points, say, $x_1$ and $x_3$. Acting on (\ref{4point}), this leads to the condition
\be
v^{\Delta} F(u,v) = u^{\Delta} F(v,u) .
\label{crossingcondition}
\ee

In solving the crossing symmetry condition (\ref{crossingcondition}), the approach that has been practiced widely is to expand the scalar function $F(u,v)$ as
\begin{equation}
F(u,v) = 1 + \sum_{\Delta, \ell}^{'} C_{\Delta,\ell}^2 \ G_{\Delta,\ell}(u,v),
\end{equation}
where the identity operator is separated from all other  operators: the summmation $\Sigma^{'}$ runs over all primary operators of nonzero scaling dimensions $\Delta \ge \Delta_{\rm min} > 0$ for zero spin. For spin $\ell$, summation contains all primary operator over unitary bound. So, an input we specify is the gap in the spectrum $\Delta_{\rm min}$. We refer this specification as one-gap bootstrapping. The crossing symmetry condition (\ref{crossingcondition}) is now recast as
\begin{align}
v^{\Delta}-u^{\Delta} &= \sum_{\Delta ,\ell}^{'} C_{\Delta,\ell}^2 \mathcal{F}(\Delta,\ell,u,v) \nonumber \\
\mathcal{F}(\Delta,\ell,u,v) &:= u^{\Delta} G_{\Delta,\ell}(v,u)-v^{\Delta} G_{\Delta,\ell}(u,v).
\label{sumrule}
\end{align}

The crossing sum rule (\ref{sumrule}) can be solved by Taylor expanding it around the symmetric point $u = v = 1/4$. Changing the variables as (\ref{uv}), solving the sum rule (\ref{sumrule}) within analytic domain of $z, \bar{z}$ amounts to solving the set of infinitely many unfolded equations at the point $z=\overline{z} = 1/2$:
\be
\mathcal{F}_0^{m,n}(\Delta,\ell,z,\bar{z}) = \sum_{\Delta,\ell}^{'} C_{\Delta,\ell}^2 \mathcal{F}^{m,n}(\Delta,\ell,z,\bar{z})
\label{sumrule2}
\ee
where
\be
\mathcal{F}^{m,n}(\Delta,\ell,z,\bar{z}) \equiv \partial_z^m \partial_{\bar{z}}^n \mathcal{F}(\Delta,\ell,z,\bar{z}) \Big|_{z=\frac{1}{2}, \bar{z}=\frac{1}{2}},
\ee
subject to boundary condition:
\be
\mathcal{F}_0 := (1-z)^{\Delta}(1-\bar{z})^{\Delta}-(z\bar{z})^{\Delta}.
\ee

The set of unfolded equtions (\ref{sumrule2}) can be solved by the linear programming \cite{Rychkov:2008a}. Define linear functional $\mathbb{L}[\cdot]$ by
\begin{equation}
\mathbb{L}[\mathcal{F}^{m,n}(\Delta,\ell,z,\bar{z})] := \sum_{m,n} \alpha_{m,n} \ \mathcal{F}^{m,n}(\Delta,\ell,z,\bar{z}),
\end{equation}
where $\alpha_{m,n}$ denotes a real coefficient. Taking this linear functional on both side of (\ref{sumrule2}),
\begin{equation}
\mathbb{L}[\mathcal{F}_0^{m,n}(\Delta,\ell,z,\bar{z})] =  \sum_{\Delta,\ell}^{'} C_{\Delta,\ell}^2 \sum_{m,n} \alpha_{m,n} \mathbb{L}[\mathcal{F}^{m,n}(\Delta,\ell,z,\bar{z})].
\label{linear_functional}
\end{equation}
We need to solve (\ref{linear_functional}) subject to the constraints that $C_{\Delta, \ell}^2$ is positive and all $\Delta$'s are above unitarity bound of respective spin($l \neq 0$). In practice, numerical method for solving (\ref{linear_functional}) requires truncation of summation  up to suitable order. In our computation below, we have done so by truncating the unfolded basis $(m,n)$ up to $m+n \equiv k \le 15$ and the spin basis $\ell$ up to $\ell \le \ell_{\rm max} = 20$.\\

%--------------------------------------------------------
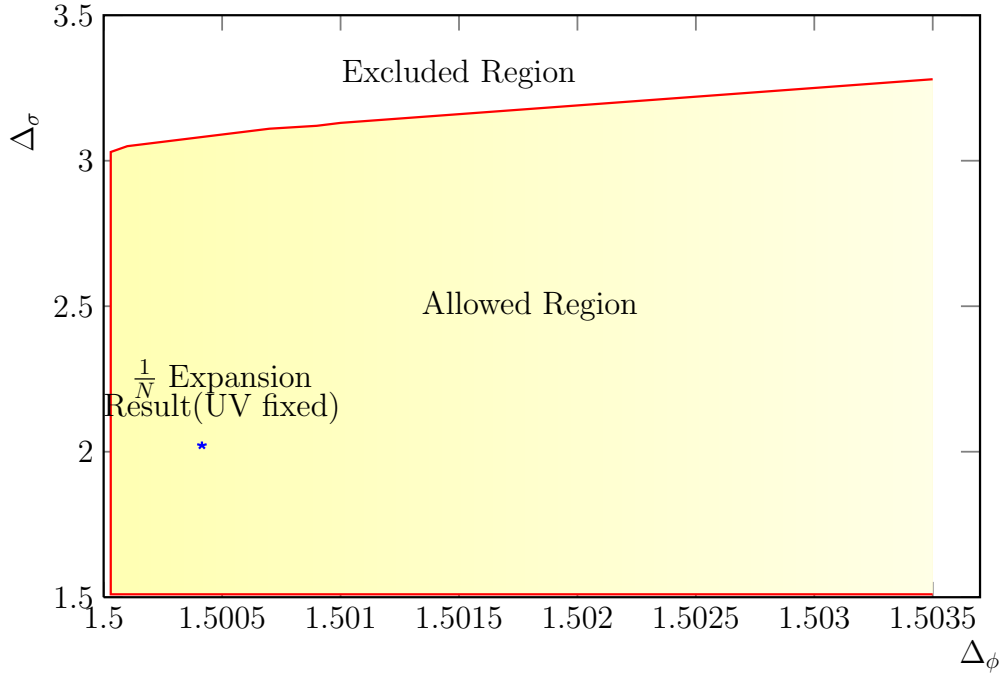
\begin{figure}
\centerline{
\begin{tikzpicture}[scale=0.85][h]
\tikzset{sha/.style={shade,left color=yellow!30!white, middle color=red, right color=yellow!10!white}}
\begin{axis}[xscale=2,yscale=1.6,
 /pgf/number format/.cd,fixed,precision=5,
%legend style={at={(0.03,0.03)},
%anchor=south west},
xlabel=$\Delta_\phi$,
ylabel=$\Delta_\sigma$,
xmin=1.500,
ymin=1.5,
xmax=1.5037,
ymax=3.5
]
\addplot [no markers,color=red,fill=yellow!30!white,sha] coordinates {
  (1.5035, 3.28) (1.503, 3.25) (1.502, 3.19)  (1.501, 3.13)  (1.5009, 3.12) (1.5008, 3.115)  (1.5007, 3.11)  (1.5006, 3.10) (1.5005, 3.09)  (1.5004, 3.08)  (1.5003, 3.07)  (1.5002, 3.06) (1.5001, 3.05)  (1.50003, 3.03)   (1.50003, 3.00)  (1.50003, 1.51) (1.5035, 1.51)
};
\addplot [mark=star,mark size=1.1pt,color=blue] coordinates {(1.5004149, 2.0215767)};
\node at (1.5,180) {Excluded Region};
\node at (1.8,100) {Allowed Region};
\node at (0.5,75) {$\frac{1}{N}$ Expansion};
\node at (0.5,65) {Result(UV fixed)};
%\legend{$d=2$,$d=3$,$d=4$,$d=5$,$d=6$}
\end{axis}
\end{tikzpicture}
}
\caption{\sl The result of one-gap numerial boostrap for $N=500$. The colored region is the values scaling dimensions consistent with the unitarity and crossing symmetry. The ultraviolet fixed point predicted by the $1/N$-expansion lies at an interior of the region.}
\end{figure}
%-------------------------------------------------------

In solving the crossing symmetry condition, an approach widely used so far assumes a single gap $\Delta_{min}$ above the unitarity bound in scalar spectrum($l=0$). The sums over $(\Delta,\ell)$ in (\ref{linear_functional}) contain all continuous operators in scalar sector($\ell=0$) but above $\Delta_{min}$ and all continuous operator in higher-spin sector($\ell \neq 0$).  If there is set of ${\alpha_{m,n}}$ that satisfy positiveness of both side of (\ref{linear_functional}) under assumption of spectrum with specific value of $\Delta_{\rm min}$, it potentially represents a conformal field theory consistent with unitarity and crossing symmetry. If not, it may still represent a conformal field theory but it must be a non-unitary one. Numerically, the unfolded conditions (\ref{linear_functional}) was solved originally in linear programming \cite{Rychkov:2008a} and later in semi-definite programming \cite{Simmons-Duffin:2011a}.

We performed the numerical bootstrap with one-gap approach. The result is shown in Figure 2. The result indicates that, in sharp contrast to the numerical bootstrap results for spacetime dimensions 4 or less, the nontrivial ultraviolet fixed point predicted by large-$N$ and $\epsilon$-expansions (which we will review in the next section) lies well below the upper boundary of the allowed region. Moreover, there is no kink structure on the upper bounary. We thus conclude that the one-gap approach does not render any specific information on nontrivial fixed point. Clearly, the one-gap approach being incapable of pinning down the critical point precisely, a better approach is sought for.

%-------------------------------------------------------
\subsection{Two-Gap Bootstrapping: Proposal}

%---------------------------------------------------------
\begin{figure}
\centerline{
  \resizebox{15.5cm}{!}{
    \begin{tikzpicture}[
      scale=1.8,
      level/.style={},
      levelthick/.style={very thick},
      virtual/.style={densely dashed},
      trans/.style={<->,shorten >=4pt,shorten <=4pt,>=stealth},
      classical/.style={thin,double,<->,shorten >=4pt,shorten <=4pt,>=stealth}
    ]
  \shade[top color=white,bottom color=gray!60!white] (0cm,9em)
    -- ++(0,-6em) -- ++(1.5cm,0) -- ++(0,6em) -- cycle;
    \draw[levelthick] (1.5cm,-2em) -- (0cm,-2em) node[pos=0,right] {
  \fontsize{8pt}{0pt} {Unit operator}};
    \draw[levelthick] (1.5cm,3em) -- (0cm,3em) node[pos=0,right] {
  \fontsize{8pt}{0pt} {Lowest operator}};
    \draw[trans] (0.5cm,-2em) -- (0.5cm,3em) node[pos=0.5,right] {
  \fontsize{15pt}{0pt} {$\Delta_{\rm{min}}$}} ;
  %%%%%%%%%%%%%%%%%%%%%%%%%%%%%%%%%%%%%%%%%%%%%%%%%
  \shade[top color=white,bottom color=gray!60!white] (5cm,9em)
    -- ++(0,-4em) -- ++(1.5cm,0) -- ++(0,4em) -- cycle;
    \draw[levelthick] (6.5cm,-2em) -- (5cm,-2em) node[pos=0,right] {
  \fontsize{8pt}{0pt} {Unit operator}};
    \draw[levelthick] (6.5cm,5em) -- (5cm,5em) node[pos=0,right] {
  \fontsize{8pt}{0pt} {Next-Lowest operator}};
    \draw[levelthick] (6.5cm,2em) -- (5cm,2em) node[pos=0,right] {
  \fontsize{8pt}{0pt} {Lowest operator}};
    \draw[trans] (6.1cm,-2em) -- (6.1cm,5em) node[pos=0.4,right] {
  \fontsize{15pt}{0pt} {$\Delta_{\rm{gap}}$}} ;
    \draw[trans] (5.4cm,-2em) -- (5.4cm,2em) node[pos=0.4,right] {
  \fontsize{15pt}{0pt} {$\Delta_{\rm{min}}$}} ;
    \end{tikzpicture}
  }
}
\caption{\sl Low-lying spectrum of one-gap approach traditionally used for $d<4$ versus two-gap approach we propose in this work. Left figure illustrates typical one-gap setup in bootstrap program. Right figure depicts our input of two-gap in the scalar operator spectrum. Above the unit operator, we have an isolated scalar operator of conformal scaling dimension $\Delta_{\rm min}$. All other operators of higher scaling dimension starts at $\Delta_{\rm gap}$. }
\label{fig_low_lying}
\end{figure}
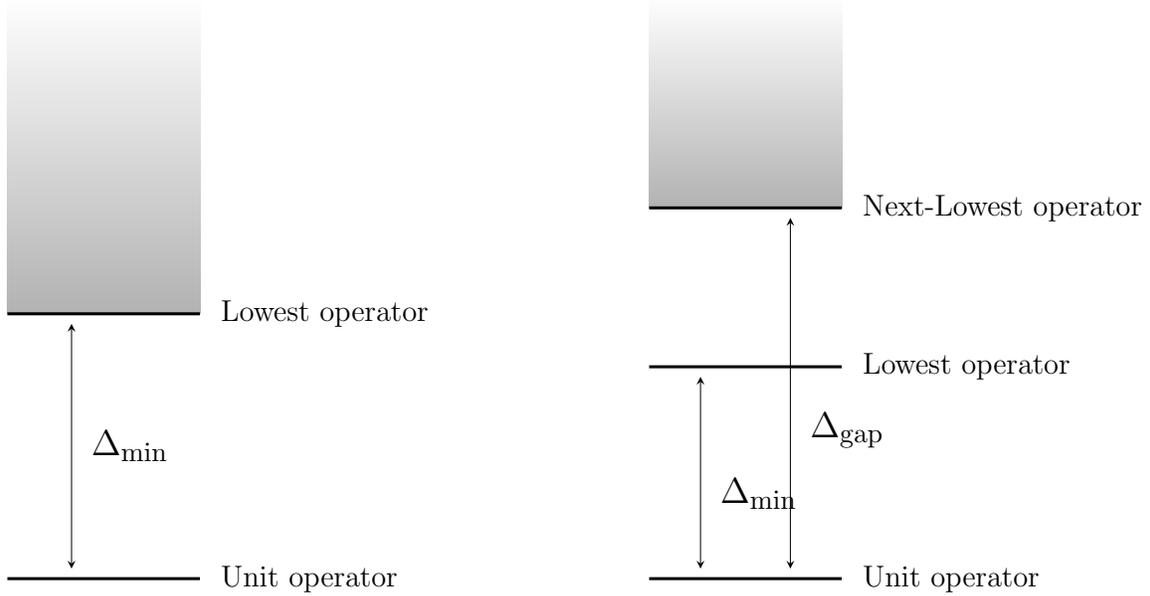
%--------------------------------------------------------

To remedy the problem alluded above that the one-gap approach is not capable of locating the ultraviolet critical point, the idea we put forward is to use two-gap approach. By this, we mean that we assume that the lowest scalar operator (other than the identity operator) has scaling dimension $\Delta_{\rm min}$ and that all other scalar operators start with scaling dimension at least $\Delta_{\rm gap}$. Our approach is most transparently depicted in Figure \ref{fig_low_lying}.

The idea is this: Compared to the one-gap bootstrapping, our two-gap bootstrapping expected to be carve out more space. This is because we are depleting primary operators in the scalar sector whose conformal scaling dimension lies between $\Delta_{\rm min}$ and $\Delta_{\rm gap}$. Suppose a potential conformal fixed point has a scaling operator in the scalar sector spectrum between $\Delta_{\rm min}$ and $\Delta_{\rm gap}$. The one-gap approach should capture this fixed point as a solution to the numerical bootstrap. On the other hand, the two-gap approach would consider this fixed point as an inconsistent theory. Therefore, we expect the two-gap approach constrains a putative conformal field theory further. A similar idea was considered in  three-dimensional bootstrap program and there it also pointed to  further restrictions for exploring ultraviolet and infrared fixed points \cite{Rychkov:2012a}. As we will see later, however, a sharp difference is that nontrivial fixed points in spacetime dimensions less than four are already located by the one-gap approach, while those in spacetime dimensions larger than four necessitates the two-gap approach at the least.

More specifically, with the $O(N)$ global symmetry at hand, the operator product of two primary scalar fields $\phi_i$ in the fundamental representation of $O(N)$ is schematically given by~\cite{Vichi:2011a}
\begin{equation}
\phi_i \times \phi_j \quad \sim \quad \sum_{S} \delta_{ij} \mathcal{O} + \sum_{T} \mathcal{O}_{(ij)} + \sum_{A} \mathcal{O}_{[ij]},
\end{equation}
where the three terms in the right-hand side refer to the singlet, symmetric traceless, and antisymmetric irreducible representation sectors, respectively. As for the spin $\ell$, the singlet and the symmetric traceless tensor sectors contain even spins only, while the antisymmetric tensor sector contain odd spins only. Reflecting this structure, sum rule for this case reads
\begin{equation}
\sum_{S,\Delta,\ell= \rm even} c_{\Delta,\ell} V_{S,\Delta,\ell} + \sum_{T,\Delta,\ell =\rm even} c_{\Delta,\ell} V_{T,\Delta,\ell} + \sum_{A,\Delta,\ell= \rm odd} c_{\Delta,\ell} V_{A,\Delta,\ell} = 0,
\end{equation}
where
%-------------------------------------------------------
\begin{align}
V_{S,\Delta,\ell} &=
\begin{pmatrix}
0 \\
\mathcal{F}^{-}_{\Delta,\ell}(u,v) \\
\mathcal{F}^{+}_{\Delta,\ell}(u,v)
\end{pmatrix}
, \quad
V_{T,\Delta,\ell}
\begin{pmatrix}
\mathcal{F}^{-}_{\Delta,\ell}(u,v) \\
(1-\frac{2}{N})\mathcal{F}^{-}_{\Delta,\ell}(u,v) \\
-(1+\frac{2}{N})\mathcal{F}^{+}_{\Delta,\ell}(u,v)
\end{pmatrix}
, \quad
V_{A,\Delta,\ell}=
\begin{pmatrix}
-\mathcal{F}^{-}_{\Delta,\ell}(u,v) \\
\ \ \mathcal{F}^{-}_{\Delta,\ell}(u,v) \\
-\mathcal{F}^{+}_{\Delta,\ell}(u,v)
\end{pmatrix} \
\nonumber
\end{align}
and
\be
\mathcal{F}^{\pm} &\equiv v^{\Delta} G_{\Delta,l}(u,v) \pm u^{\Delta} G_{\Delta,l}(v,u).
\label{O_n_sum}
\ee
In our two-gap approach, we propose to introduce two parameters ($\Delta_{\rm min}, \Delta_{\rm gap}$) into the singlet sector $V_{S,\Delta,\ell}$. For nonsinglet sectors  $V_{T,\Delta, \ell}$ and $V_{A,\Delta,\ell}$, we include all operators in so far as their scaling dimensions are above the unitary bound.

For the numerical optimization, we converted this problem into semi-definite programming~\cite{Simmons-Duffin:2011a}. We proceeded as follows. Firstly, using the radial approximation and the Zamolodchikov recursion relation (\ref{radial_app}), we expressed the function $\mathcal{F}^{m,n}(\Delta,\ell,z,\bar{z})$ as a sum over conformal blocks, in which structure of this building block is given by $\Pi_{i} \big( \frac{1}{\Delta- \Delta_i} \big) P^{m,n}_\ell(\Delta)$. This is because, successive iteration of recursion relation (\ref{radial_app}) generates product of $\frac{1}{\Delta- \Delta_i}$, which appears in (\ref{radial_app}). As $\Pi_{i} \big( \frac{1}{\Delta- \Delta_i} \big)$ is positive-definite, it suffices to focus on the  polynomial $P^{m,n}_\ell(\Delta)$.
Secondly, we parametrized scaling dimension of operators above $\Delta_{gap}$ in scalar sector by $\Delta = \Delta_{\rm gap}(1+\alpha), \ \alpha \in (0, \infty]$. This puts $P^{m,n}_\ell(\Delta=\Delta_{gap}(1+\alpha))$ a polynomial of $\alpha$.  Likewise, spin sector parametrized by $\Delta=(l+d-2)(1+\alpha), \ \alpha \in (0, \infty$]. This parametrization means we consider all operators over unitary bound. Therefore, regardless of spin, the function $P^{m,n}_\ell(\Delta)$ is essentially polynomial of $\alpha$. This polynomial structure of $P^{m,n}_\ell(\alpha)$ enables to put the optimization into semi-definite programming.

Below, we provide the pseudocode for our optimization of (\ref{linear_functional}):
\color{blue}
\vskip0.3cm
\hrule
\vskip0.3cm
{\tt Semi-Definite Programming: }
\vskip0.3cm
\hrule
\begin{align}
\tt{Minimize }& \ \ \quad \mathbb{L}[\mathcal{F}_0^{m,n}(\Delta,l,z,\bar{z})] \nonumber \\
& \hskip- 0.5cm {\tt subject \ to} \quad \mathbb{L} (P^{m,n}(\Delta_{\rm min})) >0; \nonumber \\
& \ \ \quad \vec{\alpha} = (1,\alpha,\alpha^2, \cdots, \alpha^d)
\nonumber \\
                 & \ \ \quad \mathbb{L}(P^{m,n}(\Delta_0(1+\alpha))) = \vec{\alpha}^{\rm T} \mathbb{A}_\ell \ \vec{\alpha} + \alpha(\vec{\alpha}^{\rm T}  \mathbb{B}_\ell \ \vec{\alpha})
%\quad {\tt for} \quad 0 \le \ell \le \ell_{max}
\nonumber \\
                 & \ \ \quad \Delta_0 = \Delta_{\rm gap}(1+\alpha) \quad {\tt if} \  \ \quad \ell =0 \nonumber \\
& \hskip1.8cm \Delta_{\ell}(1+\alpha) \ \quad {\tt else} \quad \ell >0 \nonumber \\
                 & \quad \ \ \mathbb{A}_\ell \succeq 0, \quad \mathbb{B}_\ell \succeq 0;  %\quad {\tt for} \quad 0 \le \ell \le \ell_{\rm max}
\nonumber \\
{\tt given} \hskip-0.7cm & \hskip1cm (N, \Delta_{\rm gap}); \nonumber \\
{\tt Run} \hskip1cm & \ell = 0 \nonumber \\
& \ell = \ell + 1 \nonumber \\
{\tt Stop} \hskip0.75cm & \ell = \ell_{\rm max}
\label{semidef}
\end{align}
\vskip0.3cm
\hrule
\vskip0.5cm
\color{black}
%------------------------------------------------------
In the code, $\Delta_\ell$ is the unitary bound for spin-$\ell$ operators, given by $(d-2+\ell)$. In our computation, we truncated the spins to $\ell \le \ell_{\rm max} = 20$. Also, $\mathbb{A}_\ell$ and $\mathbb{B}_\ell$ are matrices that built from the polynomials $P^{m,n}(\alpha)$. In our computation, we calculated the numerical value of $\mathbb{A}_\ell, \mathbb{B}_\ell$ matrix entries by {\ttfamily{Mathematica}}. These matrix entries are the input parameters of semi-definite programming. For numerical optimization of semi-definite programming with respect to the parameters $\alpha_{m,n}$, we used the open source {\ttfamily{SDPA-GMP}}.\\

%%%%%%%%%%%%%%%%%%%%%%%%%%%%%%%%%%%%%%%%%%%%%%%%%%%%%%%%%%%%%%%%%%%%%%%%%%%%%
\section{$O(N)$ Invariant Bootstrap in Five Dimensions}
\subsection{Scaling Dimensions of Light Scalar Opeartors}
The $4-\epsilon$ expansion is a widely used perturbative approach for locating the second-order phase transition and computing critical exponent. If the spacetime dimension is less than the Ginzburg criterion, $d_c = 4$,  near-critical behavior of the second-order phase transition is well described by the Landau-Ginzburg framework: their universality classes are classified by the spacetime dimensions and relevant internal symmetry. On the other hand, the second-order phase transitions in spacetime dimension above the Ginzburg criterion is described by the mean field theory, and this does not render any direct signal for possible non-trivial ultraviolet fixed point. 

To get around this difficulty and to find nontrivial ultraviolet fixed point of $O(N)$ symmetry in higher spacetime dimensions, an alternative approach based on Hubbard-Stratonovich method was considered \cite{Klebanov:2014a,Klebanov:2014b}. The theory, consisting of scalar fields ${\boldsymbol \phi}^i, \sigma$ of $O(N)$ vector and scalar representations, is defined by the Lagrangian density:
\begin{equation}
\mathcal{L} = \frac{1}{2} (\partial_m {\boldsymbol \phi^i})^2 +  \frac{1}{2} (\partial_m \sigma)^2 + {\lambda_1 \over 2} \sigma {\boldsymbol \phi^i}^2 + {\lambda_2 \over 3!} \sigma^3 \quad \quad (i=1,2,\cdots N).
\label{6d_cubic}
\end{equation}
In six-dimensional spacetime, both $\lambda_1$ and $\lambda_2$ are marginal couplings. The fixed points are classifiable by the associated $O(N)$ symmetry. Two limiting situations are of interest. If the ${\boldsymbol \phi}^i$ field becomes heavy and decoupled, the theory is reduced to a system of $O(0)$ symmetry in which the $\sigma$ scalar field dominates the dynamics with cubic self-interaction. At the fixed point, the coupling $\lambda_2$ is driven to a purely imaginary value. Therefore, this theory belongs to the universality class of the Lee-Yang edge singularity (which is a non-unitary theory). Otherwise, the $O(N)$ vector field ${\boldsymbol \phi}^i$ couples to a system of the $\sigma$ scalar field with bosonic Yukawa-type interactions. Starting from the Gaussian fixed point, there would be the renormalization group flows leading to these fixed points. 

The perturbative computation of this system in $1/N$ and $\epsilon$ double expansion was performed in \cite{Klebanov:2014a,Klebanov:2014b}. Their result indicates that both situations of the fixed point is captured as $N$ is varied: fixed point values of the coupling constants $\lambda_1, \lambda_2$ are real-valued for sufficiently large $N$, while complex-valued for sufficiently small $N$. The spacetime dimension is above the Ginzburg criterion, so the flow between ultraviolet fixed point and the infrared fixed point is reversed compared to the spacetime dimension less than four.  

In five-dimensional spacetime, the scaling dimensions of ${\boldsymbol \phi}^i$ and $\sigma$ are computable perturbatively. They were computed up to third orders in $1/N$-expansion \cite{Klebanov:2014a,Kotikov:1996a}. The result is
\begin{align}
\Delta_{\boldsymbol \phi} &= \frac{3}{2} + \frac{0.216152}{N} - \frac{4.342}{N^2} - \frac{121.673}{N^3} + \cdots \nonumber \\
\Delta_\sigma &= 2+ \frac{10.3753}{N} + \frac{206.542}{N^2} + \cdots
\label{anomalous_dimension}
\end{align}
For sufficiently large $N$, we expect the critical theory to exist at $(\Delta_\phi, \Delta_\sigma)=(\frac{3}{2},2)$, distinguished from the free theory at $(\Delta_\phi, \Delta_\sigma)=(\frac{3}{2},3)$. However, the above perturbtive result indicates that, for sufficiently small $N$, negative contribution of $\frac{1}{N}$ corrections dominate. In this case, $\Delta_\phi$ falls below the unitary bound $3/2$ of five-dimensional scalar operator. This suggests that, at sufficiently small $N$, the critical fixed point should be interpreted as describing a non-unitary theory.

%To better approach the existence of critical fixed point and possible non-unitarity at sufficiently low $N$, we use the conformal bootstrap program.   Bootstrap program can covers non-perturbative area, therefore it will give some information on exact behavior of UV fixed point.

%---------------------------------------------------------
\subsection{Bootstrap Results}
We considered the $O(N)$ global symmetric bootstrap, where the sum rule was decomposed according to (\ref{O_n_sum}).
We carried out the numerical bootstrapping with the proposed two-gap approach in the scalar sector $V_{S,\Delta,\ell}$ by semi-definite programming.
We identified regions in $(\Delta_{\boldsymbol \phi}, \Delta_\sigma)$ space where the unitarity and the crossing symmetry conditions are satisfied. We repeated the procedure with varying $N, \Delta_{\rm min}, \Delta_{\rm gap}$ and addressed the following questions.

%--------------------------------------------------------
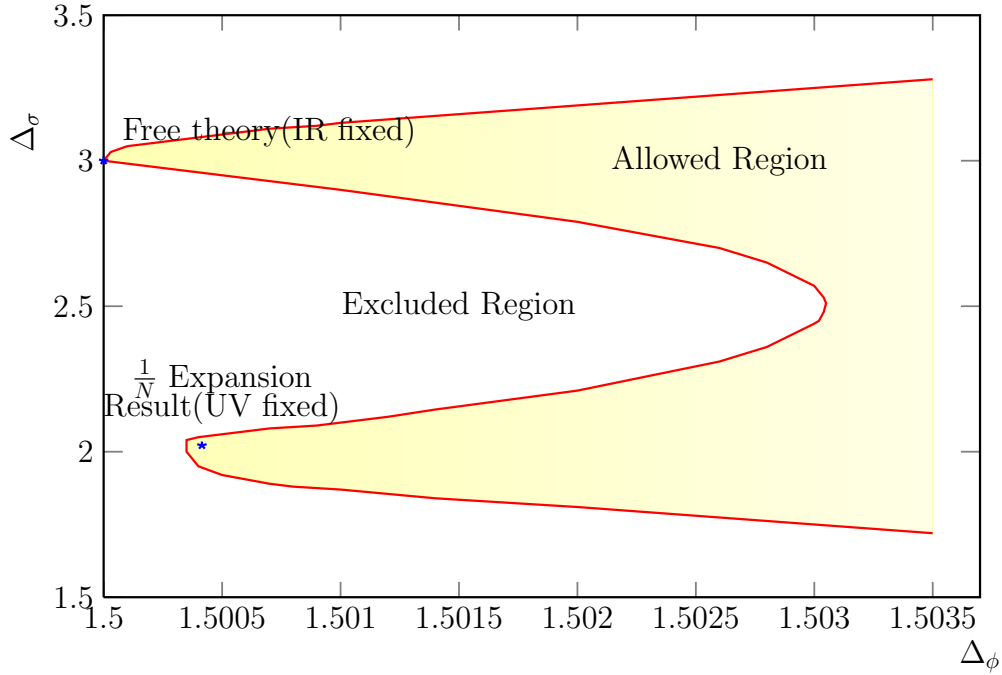
\begin{figure}

\centerline{
\begin{tikzpicture}[scale=0.85][h]
\tikzset{sha/.style={shade,left color=yellow!30!white, middle color=red, right color=yellow!10!white}}
\begin{axis}[xscale=2,yscale=1.6,
 /pgf/number format/.cd,fixed,precision=5,
%legend style={at={(0.03,0.03)},
%anchor=south west},
xlabel=$\Delta_\phi$,
ylabel=$\Delta_\sigma$,
xmin=1.500,
ymin=1.5,
xmax=1.5037,
ymax=3.5
]
\addplot [no markers,color=red,fill=yellow,sha] coordinates {
(1.5035, 1.72) (1.503, 1.75) (1.502, 1.81) (1.5014, 1.84)  (1.5012 , 1.855)  (1.501, 1.87) (1.5009 , 1.875)  (1.5008 , 1.88)  (1.5007 , 1.89)  (1.5006, 1.905)  (1.5005, 1.92) (1.5004, 1.95) (1.50035, 2.00)  (1.50035, 2.04)  (1.5004, 2.05)  (1.5005, 2.06)  (1.5006, 2.07)  (1.5007, 2.08)  (1.5008, 2.085)  (1.5009, 2.09)  (1.501, 2.10)  (1.5012, 2.12)  (1.5014, 2.145)  (1.502, 2.21)  (1.5026, 2.31)  (1.5028, 2.36)  (1.503, 2.44)  (1.50302, 2.45)  (1.50304, 2.48)  (1.50305, 2.51) (1.50305 , 2.51)  (1.50304, 2.53)  (1.50302, 2.55)  (1.503, 2.57)  (1.5028, 2.65)  (1.5026, 2.70)  (1.502, 2.79)  (1.501, 2.90)  (1.5009, 2.91)  (1.5008, 2.92)  (1.5007, 2.93)  (1.5006, 2.94)  (1.5005, 2.95)  (1.5004, 2.96)  (1.5003, 2.97)  (1.5002, 2.98)  (1.5001, 2.99)  (1.5000, 3.00)  (1.50003, 3.03)  (1.5001, 3.05)  (1.5002, 3.06)  (1.5003, 3.07)  (1.5004, 3.08)  (1.5005, 3.09)  (1.5006, 3.10)  (1.5007, 3.11)  (1.5008, 3.115)  (1.5009, 3.12)  (1.501, 3.13)  (1.502, 3.19)  (1.503, 3.25)  (1.5035, 3.28)
};
\addplot [mark=star,mark size=1.1pt,color=blue] coordinates {(1.5004149, 2.0215767)};
\addplot [mark=star,mark size=1.1pt,color=blue] coordinates {(1.50, 3.00)};
\node at (1.5,100) {Excluded Region};
\node at (2.6,150) {Allowed Region};
\node at (0.7,160) {Free theory(IR fixed)};
\node at (0.5,75) {$\frac{1}{N}$ Expansion};
\node at (0.5,65) {Result(UV fixed)};
%\legend{$d=2$,$d=3$,$d=4$,$d=5$,$d=6$}
\end{axis}
\end{tikzpicture}
}
\caption{\sl Result of two-gap approach for $N=500$ and $k=15$. Yellow-colored part is the allowed region, consistent with the unitarity and the crossing symmetry of 4-point correlation function. Compared to the one-gap approach result in Figure 2, the two-gap approach carves out regions of low values above the unitarity bound. Its boundary features two cusps. The ultraviolet nontrivial fixed point is located at its lower tip, while the infrared Gaussian fixed point is located at its upper tip.}
\label{N_500_result}
\end{figure}

\begin{itemize}
\item Does the two-gap approach constrain the theory space  more than the one-gap approach? Is the two-gap approach enough to locate both the Gaussian and nontrivial fixed points on its allowed region boundary?
\item What is the range of validity of perturbative $1/N$-expansions?
\item How do locations of the fixed points move around as the theory parameters $N, \Delta_{\rm gap}$ are varied?
\item At extreme values of $N, \Delta_{\rm gap}$, do fixed points appear or disappear? If so, what are critical value $N^{\rm crit}, \Delta^{\rm crit}_{\rm gap}$ for onset of such behavior?
\item From scaling consideration, we expect that bootstrapping for $d>4$ and bootstrapping for $d<4$ are dual each other in that ultraviolet and infrared regimes are interchanged. Do we find such `duality' from the result?
\end{itemize}

%---------------------------------------------------------
\subsubsection{Carving Out}

We first explore whether the two-gap approach curves out regions that were allowed within the one-gap approach.
For the representative choice of $N=500$, the result is shown in Figure \ref{N_500_result}.

%--------------------------------------------------------
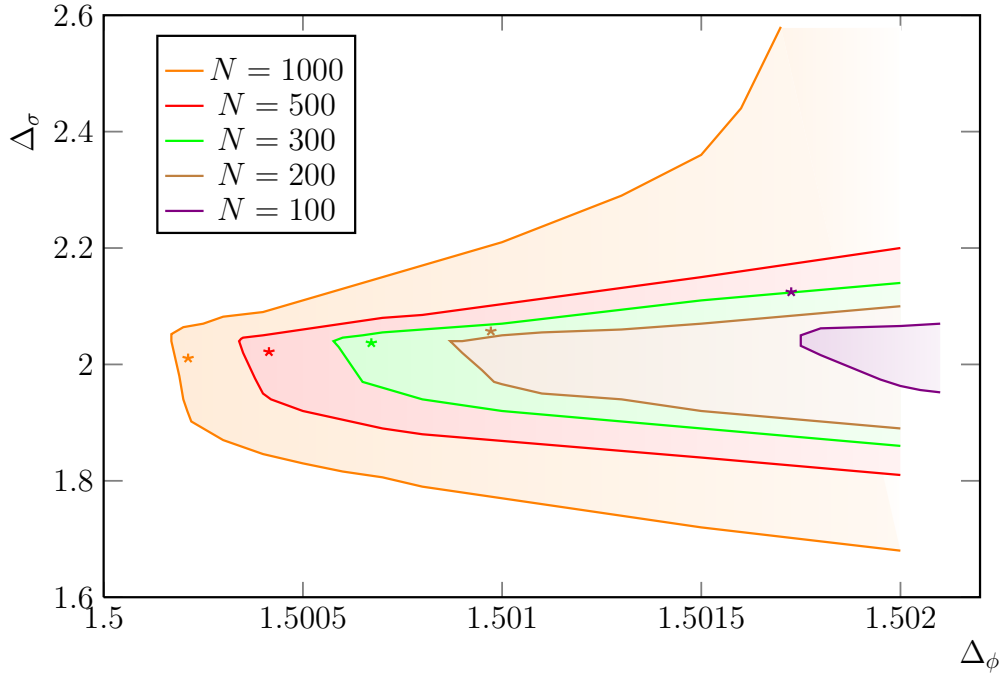
\begin{figure}
\centerline{
\begin{tikzpicture}[scale=0.85][h]
\tikzset{sha/.style={shade,left color=violet!15!white, middle color=red, right color=violet!5!white}}
\tikzset{sha2/.style={shade,left color=brown!15!white, middle color=red, right color=brown!5!white}}
\tikzset{sha3/.style={shade,left color=green!15!white, middle color=red, right color=green!5!white}}
\tikzset{sha4/.style={shade,left color=red!15!white, middle color=red, right color=red!5!white}}
\tikzset{sha5/.style={shade,left color=orange!15!white, middle color=red, right color=orange!5!white}}
\tikzset{sha6/.style={shade,left color=orange!8!white, middle color=red, right color=orange!1!white}}
\begin{axis}[xscale=2,yscale=1.6,
 /pgf/number format/.cd,fixed,precision=5,
legend style={at={(0.03,0.39)},
anchor=south west},
xlabel=$\Delta_\phi$,
ylabel=$\Delta_\sigma$,
xmin=1.500,
ymin=1.6,
xmax=1.5022,
ymax=2.6
]
\addplot [no markers,draw=white,sha6] coordinates {(1.5017, 2.58)(1.502, 2.58)(1.502, 1.68)};
\addlegendentry{}
\addplot [no markers,color=orange,sha5] coordinates {(1.5017, 2.58) (1.5016, 2.44)(1.5015, 2.36) (1.5013, 2.29) (1.501, 2.21)(1.5006, 2.13) (1.5004, 2.09) (1.5003, 2.082) (1.50025, 2.07) (1.5002, 2.064) (1.50019, 2.06) (1.50018, 2.056) (1.500175, 2.054) (1.50017, 2.052) (1.50017, 2.04) (1.500175, 2.025) (1.50018, 2.01) (1.50019, 1.98) (1.5002, 1.94) (1.50021, 1.92) (1.50022, 1.902) (1.50025, 1.89) (1.5003, 1.87) (1.5004, 1.846) (1.5005, 1.83) (1.5006, 1.816) (1.5007, 1.806) (1.5008, 1.79)(1.501, 1.77)(1.5015, 1.72)(1.502, 1.68)};
\addlegendentry{$N=1000$}
\addplot [no markers,color=red,sha4] coordinates {(1.502, 2.20) (1.5015, 2.15) (1.5008, 2.085) (1.5007, 2.08) (1.5006, 2.07) (1.5005, 2.06) (1.5004, 2.05) (1.50035, 2.046) (1.500345, 2.044) (1.50034, 2.04) (1.500345, 2.03) (1.50035, 2.02) (1.50038, 1.976) (1.5004, 1.95) (1.50042, 1.94) (1.5005, 1.92) (1.5006, 1.905) (1.5007, 1.89) (1.5008, 1.88) (1.5015, 1.84)  (1.502, 1.81)};
\addlegendentry{$N=500$}
\addplot [no markers,color=green,sha3] coordinates {(1.502, 2.14) (1.5015, 2.11) (1.501, 2.07) (1.5008, 2.06) (1.5007, 2.055) (1.50065, 2.05) (1.5006, 2.046) (1.500577, 2.04) (1.50059, 2.03) (1.5006, 2.02) (1.50063, 1.99) (1.50065, 1.97) (1.5007, 1.96) (1.5008, 1.94) (1.501, 1.92) (1.5015, 1.89) (1.502, 1.86)};
\addlegendentry{$N=300$}
\addplot [no markers,color=brown,sha2] coordinates {(1.502, 1.89) (1.5015, 1.92) (1.5013, 1.94)  (1.5011, 1.95) (1.5010, 1.966) (1.50098, 1.97) (1.50095, 1.99) (1.5009, 2.02) (1.50087, 2.04) (1.5009, 2.04) (1.5010, 2.05) (1.5011, 2.055)(1.5013, 2.06)(1.5015, 2.07)(1.502, 2.10)};
\addlegendentry{$N=200$}
\addplot [no markers,color=violet,sha] coordinates {(1.5021, 2.07) (1.502, 2.066) (1.5019, 2.064) (1.5018, 2.062) (1.50175, 2.05) (1.50175, 2.032)  (1.5018, 2.016)  (1.50185, 2.002) (1.5019, 1.988) (1.50195, 1.974) (1.502, 1.963) (1.50205, 1.956) (1.5021, 1.952)};
\addlegendentry{$N=100$}
\addplot [mark=star,mark size=1.3pt,color=violet,scale=0.1] coordinates {(1.501726, 2.124407)};
\addplot [mark=star,mark size=1.3pt,color=brown,scale=0.1] coordinates {(1.500972, 2.057040)};
\addplot [mark=star,mark size=1.3pt,color=green,scale=0.1] coordinates {(1.500672, 2.036879)};
\addplot [mark=star,mark size=1.3pt,color=red,scale=0.1] coordinates {(1.500415, 2.02158)};
\addplot [mark=star,mark size=1.3pt,color=orange,scale=0.1] coordinates {(1.500212, 2.010582)};
%\node at (1.3,400) {Consistent Region};
%\node at (0.7,220) {Excluded Region};
%\legend{$N=100$, $N=200$, $N=300$, $N=500$, $N=1000$}
\end{axis}
\end{tikzpicture}
}
\caption{\sl Result for $\Delta_{\rm{gap}}=8.0$. Here we zoomed in around near lower tip. From leftmost, each bound stands for $N=1000, N=500, N=300, N=200, N=100$, respectively. The star marks indicate location of perturbative $\frac{1}{N}$ expansion result for each $N$. For sufficiently large $N$, star mark location gradually approaches to boundary of allowed region.}
\label{gap800a}
\end{figure}
%--------------------------------------------------------

We already presented physical reason why we expect the two-gap approach puts more restrictive result than one-gap approach. Indeed, the two-gap approach result in Figure \ref{N_500_result} further carves out the region allowed  by the one-gap approach. In this result, we have set $\Delta_{gap}$ to 8.00 and $k=15$. The result manifests two pronounced tips. The apex of upper tip region indicates free theory. Surprisingly, end of lower tip is quiet close to perturbation result (\ref{anomalous_dimension}).

Next, we bootstrapped with various $N$, fixing parameter $\Delta_{gap}=8.0$ as before. To see agreement of UV fixed point and shrapened end of tip, we zoomed in near low-tip area. The result displayed in Figure \ref{gap800a}. Each star mark is perturbation result from (\ref{anomalous_dimension}). Perturbative result of $\frac{1}{N}$ expansion for $N=100,200$ lies outside of allowed region. Our result shows bootstrap result and large-$N$ expansion are comparable when $N$ is larger than 300. Another notable point here is appearance of kink at lower bound. For $N=100$, kink do not appears while other case shows sudden change of slope.

%-------------------------------------------------------
\begin{figure}
\centerline{
\begin{tikzpicture}[scale=0.85][h]
\tikzset{sha/.style={shade,left color=red!6!white, right color=cyan!3!white}}
\tikzset{sha2/.style={shade,left color=orange!15!white, middle color=red, right color=orange!5!white}}
\tikzset{sha3/.style={shade,left color=red!15!white, middle color=red, right color=red!5!white}}
\tikzset{sha4/.style={shade,left color=cyan!4!white, middle color=red, right color=cyan!1!white}}
\begin{axis}[xscale=2,yscale=1.6,
 /pgf/number format/.cd,fixed,precision=5,
legend style={at={(0.03,0.44)},
anchor=south west},
xlabel=$\Delta_\phi$,
ylabel=$\Delta_\sigma$,
xmin=1.500,
ymin=1.6,
xmax=1.5021,
ymax=2.55
]
\addplot [no markers,draw=white, line width=1pt,sha4] coordinates {(1.5008, 1.64)(1.502, 1.64)(1.502, 2.49)(1.50144, 2.49)};
\addlegendentry{}
\addplot [no markers,color=blue, line width=1pt,sha] coordinates {(1.50144, 2.49) (1.5014, 2.44) (1.50135, 2.39)(1.5013, 2.36) (1.501, 2.25) (1.5007, 2.17) (1.5006, 2.14) (1.5005, 2.12) (1.5004, 2.1) (1.5002, 2.05) (1.50013, 2.04) (1.50012, 2.038) (1.50011, 2.034) (1.5001, 2.032) (1.50009, 2.03) (1.50009, 2.01) (1.500095, 1.97) (1.5001, 1.93) (1.50011, 1.86) (1.50012, 1.83) (1.50013, 1.82) (1.5004, 1.728) (1.5006, 1.69) (1.5008, 1.64)};
\addlegendentry{$N=2000$}
\addplot [solid,no markers,color=orange, line width=1pt,sha2] coordinates {(1.502, 2.22) (1.5015, 2.16) (1.501, 2.11) (1.5006, 2.07) (1.5004, 2.05) (1.50025, 2.034) (1.50022, 2.032) (1.5002, 2.03) (1.50019, 2.028) (1.50018, 2.026) (1.50018, 2.01) (1.50019, 1.98) (1.5002, 1.94) (1.50021, 1.92) (1.50022, 1.902) (1.50025, 1.89) (1.5003, 1.87) (1.5004, 1.846) (1.5005, 1.83) (1.5006, 1.816) (1.5007, 1.806) (1.5008, 1.79)(1.501, 1.77)(1.5015, 1.72)(1.502, 1.68)};
\addlegendentry{$N=1000$}
\addplot [no markers,color=red,sha,line width=1pt,sha3] coordinates {(1.502, 2.10) (1.5015, 2.08) (1.501, 2.05) (1.5006, 2.036) (1.5005, 2.032) (1.5004, 2.026) (1.50038, 2.02) (1.50037, 2.00) (1.50038, 1.976) (1.5004, 1.95) (1.50042, 1.94) (1.5005, 1.92) (1.5006, 1.905) (1.5007, 1.89) (1.5008, 1.88) (1.501, 1.87) (1.5015, 1.84)  (1.502, 1.81)};
\addlegendentry{$N=500$}
\addplot [mark=star,mark size=1.3pt,color=orange,scale=0.1] coordinates {(1.500415, 2.021577)};
\addplot [mark=star,mark size=1.3pt,color=red,scale=0.1] coordinates {(1.5002118, 2.0105818)};
\addplot [mark=star,mark size=1.3pt,color=blue,scale=0.1] coordinates {(1.50010699, 2.0052393)};
%\node at (1.3,400) {Consistent Region};
%\node at (0.7,220) {Excluded Region};
%\legend{$N=500$, $N=1000$, $N=2000$}
\end{axis}
\end{tikzpicture}
}
\caption{\rm Result for $\Delta_{\rm{gap}}=40.0$. Here, we zoomed in near lower tip. From leftmost, each bound presents $N=2000, N=1000, N=500$, respectively. Each star mark is the location of perturbative result from (\ref{anomalous_dimension}). The endpoint of tip agrees to star mark, identified as the ultraviolet fixed point.}
\label{gap4000}
\end{figure}
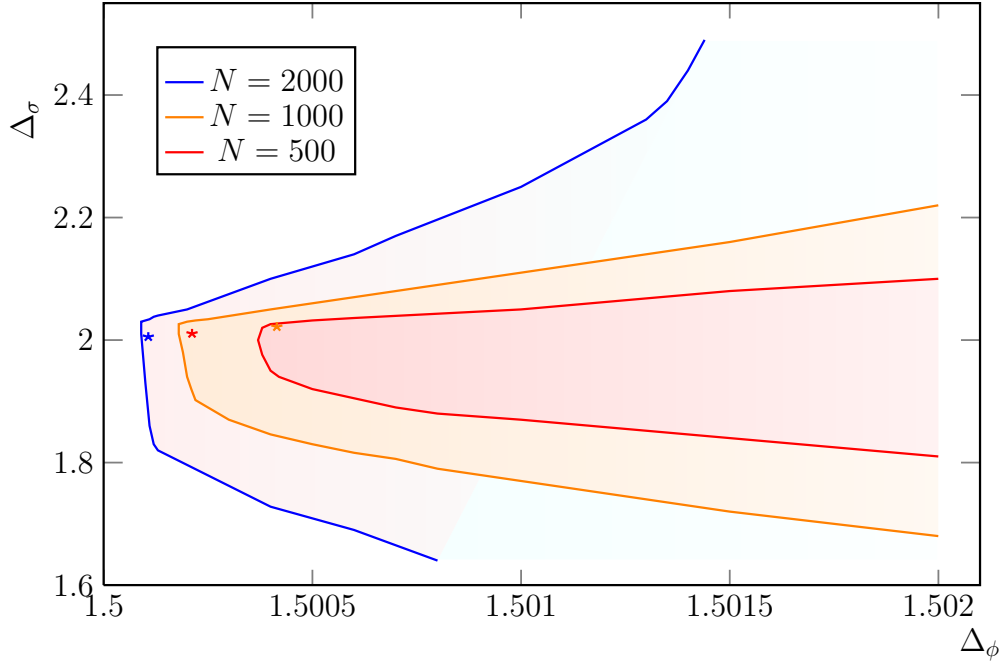
%-------------------------------------------------------

\subsubsection{Varying $\Delta_{\rm gap}$}

We also examined the impact of varying the parameter $\Delta_{\rm gap}$ at a fixed value of $N$. We increased it to $\Delta_{\rm gap}=40.0$, which is much larger than the value we set in Figure \ref{gap800a}. Since we are now ruling out more theories than for $\Delta_{gap}=8.0$, we expect the result carving more space out. The result displayed in Figure \ref{gap4000} indeed demonstrate that our intuition is met.

The pattern of carving out is worth of nothing. Overall, the boundary curve of the allowed region retains the shape of $\Sigma$. As the gap $\Delta_{gap}$ is increased, the depletion mines out and pushes the mid-part of the boundary curve (the part that takes $\Large{>}$-shape) to the right. On ther other hand, the outer boundaries -- the upper boundary emanating from the upper tip and lower boundary emanating from the lower tip - are little changed. We confirmed that this behavior persists even if   $\Delta_{\rm gap}$ is increased up to 100.0.

%--------------------------------------------------------
\subsubsection{Validity of Large-$N$ Expansion}
We also checked validity of the $1/N$-expansion. Even if the band gap is large, as in Figure \ref{gap4000} with $\Delta_{\rm gap} = 40.0$, the ultraviolet nontrivial fixed point predicted by $1/N$ expansion (note that this expansion comes with large coefficients) sits close to the tip of the allowed region. This is the behavior we already observed for lower value of $\Delta_{\rm gap}$, as in   Figure \ref{N_500_result} for $\Delta_{\rm gap} = 8.0$. %perturbation result of $N=500, 1000, 2000$ shows agreement to end point of tip.

From the proximity of the pertubative fixed point to the boundary of allowed region, we also draw a conclusion that the $1/N$-expansion becomes less reliable at larger band gap $\Delta_{\rm gap}$. This can be gleaned from the data for $N=500$. For $\Delta_{\rm gap} = 8.0$, Figure {\ref{gap800a} indicates the perturbative fixed point was enclosed by the boundary curve. On the other hand, for $\Delta_{\rm gap} = 40.0$, Figure \ref{gap4000} indicates the perturbative fixed point hits the boundary curve. Inferred from Figure \ref{gap800a} to the trend of varying $N$, it is expected that the perturbative fixed point will lies outside the allowed region for $N$ less than 500.

We consider the large-$N$ match in $(\Delta_{\boldsymbol \phi}, \Delta_{\rm min})$-space between the nontrivial fixed point predicted by large-$N$ expansion and the tip of allowed $\Sigma$-region is a strong indication that the two-gap approach is a useful method for locating nontrivial fixed point at any $N$.

%-------------------------------------------------------
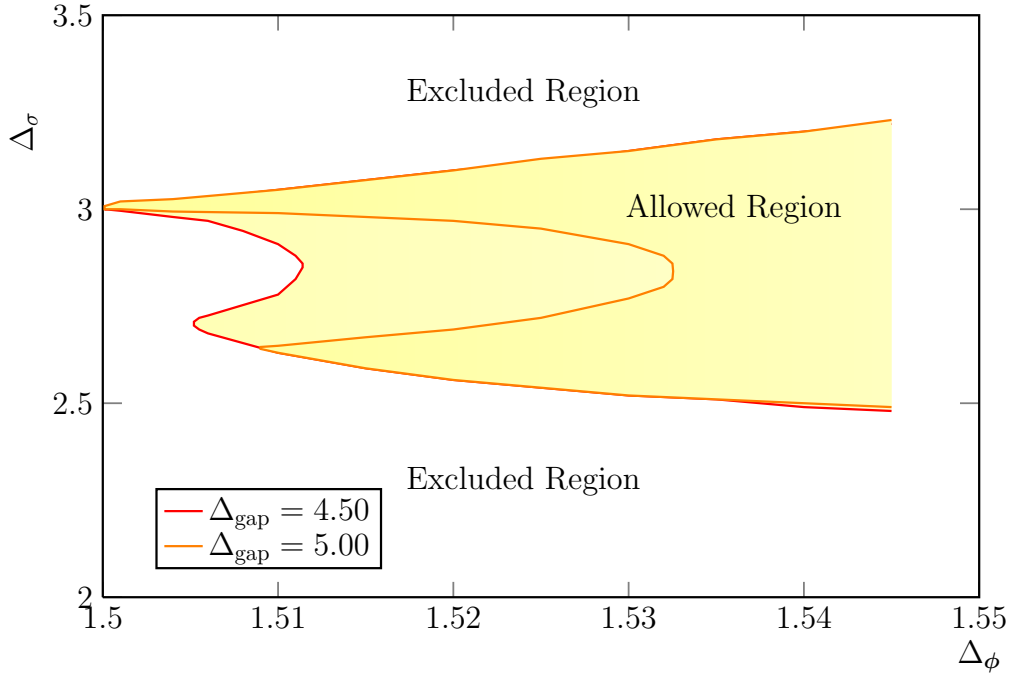
\begin{figure}
\centerline{
\begin{tikzpicture}[scale=0.85][h]
\tikzset{sha/.style={shade,left color=yellow!30!white, middle color=red, right color=yellow!20!white}}
\tikzset{sha2/.style={shade,left color=yellow!40!white, middle color=red, right color=yellow!25!white}}
\begin{axis}[xscale=2,yscale=1.6,
 /pgf/number format/.cd,fixed,precision=5,
legend style={at={(0.03,0.03)},
anchor=south west},
xlabel=$\Delta_{\boldsymbol \phi}$,
ylabel=$\Delta_\sigma$,
xmin=1.500,
ymin=2.0,
xmax=1.55,
ymax=3.5
]
\addplot [no markers,color=red,sha] coordinates {
(1.545, 3.22) (1.54, 3.2) (1.535, 3.18) (1.53, 3.15) (1.525, 3.12) (1.52, 3.1) (1.51, 3.05) (1.506, 3.03) (1.504, 3.02) (1.501, 3.004) (1.5001, 3.) (1.5001, 3.) (1.501, 2.996) (1.504, 2.98) (1.506, 2.97) (1.508, 2.944) (1.51, 2.91) (1.511, 2.88) (1.5114, 2.86) (1.5114, 2.85) (1.511, 2.82) (1.51, 2.78) (1.506, 2.726) (1.5055, 2.72) (1.5052, 2.71) (1.5052, 2.7) (1.5055, 2.69) (1.506, 2.68) (1.51, 2.63) (1.515, 2.59) (1.52, 2.56) (1.525, 2.54) (1.53, 2.52) (1.535, 2.51) (1.54, 2.49) (1.545, 2.48)
};
\addplot [no markers,color=orange,sha2] coordinates {
(1.545, 3.23) (1.54, 3.2) (1.535, 3.18) (1.53, 3.15) (1.525, 3.13) (1.52, 3.1) (1.51, 3.05) (1.504, 3.026) (1.501, 3.02) (1.5001, 3.008) (1.5001, 3.) (1.501, 3.) (1.504, 2.994) (1.51, 2.99) (1.52, 2.97) (1.525, 2.95) (1.53, 2.91) (1.532, 2.88) (1.5325, 2.86) (1.53254, 2.84) (1.53254, 2.84) (1.5325, 2.82) (1.532, 2.8) (1.53, 2.77) (1.525, 2.72) (1.52, 2.69) (1.515, 2.67) (1.51, 2.648) (1.509, 2.645) (1.509, 2.64) (1.51, 2.63) (1.515, 2.59) (1.52, 2.56) (1.525, 2.54) (1.53, 2.52) (1.535, 2.51) (1.54, 2.5) (1.545, 2.49)
};
\node at (24,30) {Excluded Region};
\node at (36,100) {Allowed Region};
\node at (24,130) {Excluded Region};
\legend{$\Delta_{\rm{gap}}=4.50$,$\Delta_{\rm{gap}}=5.00$}
\end{axis}
\end{tikzpicture}
}
\caption{\sl Result for $N=1$ and $\Delta_{\rm{gap}}=4.5, \ 5.0$. The plot features two tips, one ending at the free field theory location $(1.5, 3.0)$. In the spirit of the large $N$ counterparts, we identify the lower tip as the candidate for a nontrivial ultraviolet fixed point. Its location is sensitive to the specification of $\Delta_{gap}$.}
\label{N_1_result}
\end{figure}
%---------------------------------------------------------
\begin{figure}
\centerline{
\begin{tikzpicture}[scale=0.85][h]
\tikzset{sha/.style={shade,left color=yellow!30!white, middle color=red, right color=yellow!20!white}}
\begin{axis}[xscale=2,yscale=1.6,
 /pgf/number format/.cd,fixed,precision=5,
legend style={at={(0.03,0.03)},
anchor=south west},
xlabel=$\Delta_{\boldsymbol \phi}$,
ylabel=$\Delta_\sigma$,
xmin=1.500,
ymin=2.0,
xmax=1.940,
ymax=5.0
]
\addplot [no markers,color=red,sha] coordinates {
(1.9, 4.49) (1.834, 4.32) (1.7, 3.84) (1.65, 3.65) (1.6, 3.45) (1.57, 3.33) (1.55, 3.24) (1.53, 3.15) (1.52, 3.105) (1.51, 3.054) (1.504, 3.022) (1.504, 3.02) (1.51, 3.048) (1.52, 3.088) (1.53, 3.13) (1.55, 3.19) (1.57, 3.24) (1.6, 3.32) (1.65, 3.42) (1.7, 3.48) (1.75, 3.51) (1.8, 3.46) (1.82, 3.4) (1.834, 3.35) (1.84, 3.27) (1.842, 3.24) (1.844, 3.16) (1.844, 3.16) (1.842, 3.09) (1.84, 3.05) (1.834, 2.99) (1.8, 2.81) (1.78, 2.75) (1.77, 2.72) (1.75, 2.67) (1.72, 2.61) (1.7, 2.58) (1.65, 2.52) (1.6, 2.47) (1.58, 2.45) (1.58, 2.44) (1.6, 2.43) (1.65, 2.4) (1.7, 2.4) (1.72, 2.41) (1.75, 2.4) (1.78, 2.41) (1.8, 2.42) (1.834, 2.43) (1.9, 2.47)
};
\node at (340,170) {Allowed Region};
\node at (90,220) {Excluded Region};
\end{axis}
\end{tikzpicture}
}
\caption{\sl Result for $N=1$ and $\Delta_{\rm{gap}}=7.0$. The plot still features two tips, but tip locations reveal theory's sensitivity to the specification of $\Delta_{\rm gap}$. In particular, not only the lower tip (identified with nontrivial ultraviolet fixed point) location changes considerably, the upper tip is detached from the free field theory location. The larger $\Delta_{\rm{gap}}$ is, the more the upper tip deviates from this location.
}
\label{N_1_result}
\end{figure}
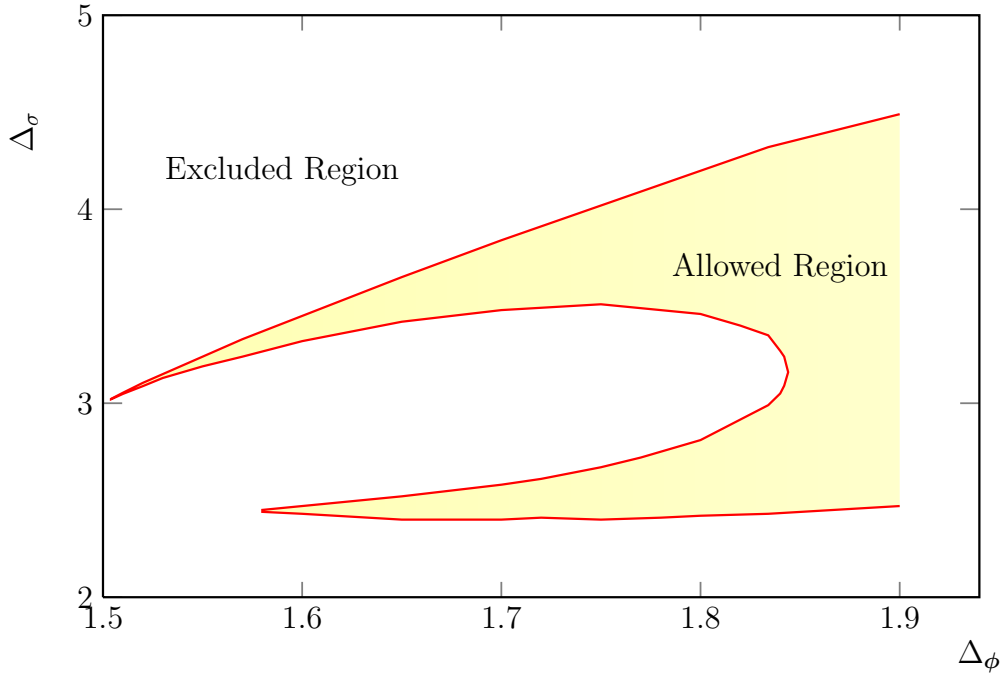
%--------------------------------------------------------

%--------------------------------------------------------
%\bigskip
\begin{figure}
\centerline{
\begin{tikzpicture}[scale=0.85][h]
\tikzset{sha/.style={shade,left color=yellow!30!white, middle color=red, right color=yellow!20!white}}
\begin{axis}[xscale=2,yscale=1.6,
 /pgf/number format/.cd,fixed,precision=5,
legend style={at={(0.03,0.03)},
anchor=south west},
xlabel=$\Delta_\phi$,
ylabel=$\Delta_\sigma$,
xmin=1.500,
ymin=2.0,
xmax=2.700,
ymax=3.5
]
\addplot [no markers,color=red,sha] coordinates {
(2.6, 3.26) (2.5, 3.11) (2.48, 3.09) (2.45, 3.04) (2.44, 3.026) (2.43, 3.01) (2.41, 2.98) (2.38, 2.94) (2.34, 2.88) (2.32, 2.85) (2.3, 2.83) (2.3, 2.83) (2.32, 2.84) (2.34, 2.86) (2.38, 2.9) (2.41, 2.93) (2.43, 2.95) (2.44, 2.964) (2.45, 2.974) (2.48, 3.003) (2.5, 3.03) (2.6, 3.14)
};
\addplot [mark=star,mark size=1.6pt,color=black,scale=0.1] coordinates {(1.500, 3.00)};
\node at (12,104) {Free Theory};
\end{axis}
\end{tikzpicture}
}
\caption{\sl Bootstrap result for $N=1$ at $k=15$. In this plot, our setup is $k=15$ and $\Delta_{\rm{gap}}=500.0$. We only focused on near-upper tip area. This upper sharpened point now far from free theory. Increasing $\Delta_{\rm{gap}}$ into 700, 1000 gives same result, increasing $\Delta_{\rm{gap}}$ has limit. Therefore, if there exist a theory such that finite operator contents on scalar sector, fixed point of this theory would not be described by free theory.}
\label{N_1_result-2}
\end{figure}
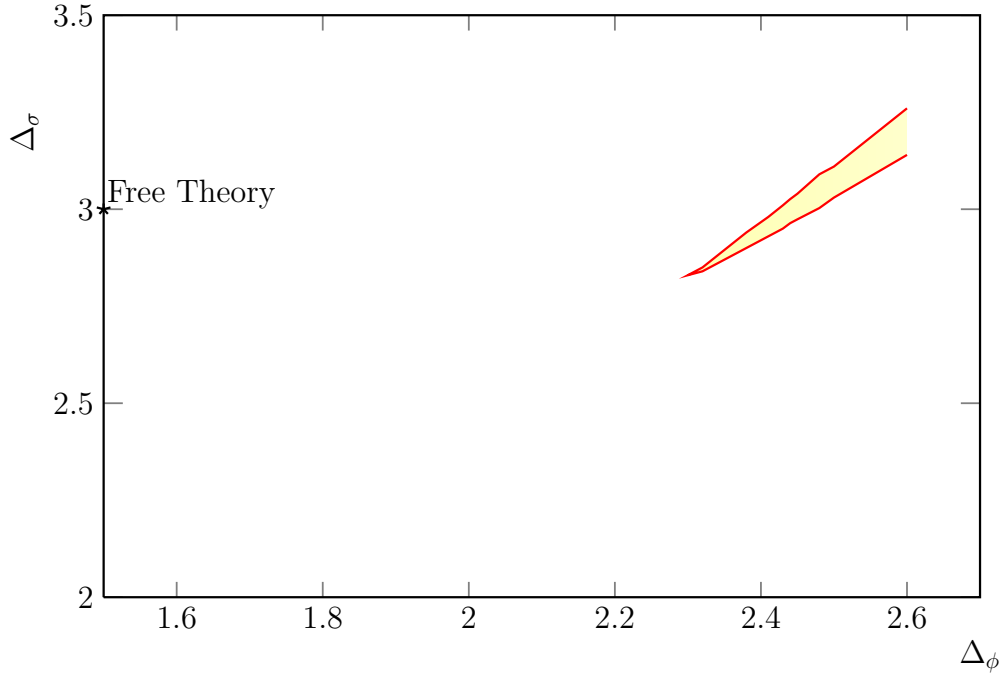
%--------------------------------------------------------

%--------------------------------------------------------
\subsubsection{$N = 1$ versus (Non)unitarity}
The perturbative $\frac{1}{N}$-expansion result (\ref{anomalous_dimension}) yields imaginary value of the fixed point coupling constants at small values of $N$, suggesting breakdown of the unitarity. We thus explored signals of unitarity breakdown.

To take the extreme, we took the smallest possible value $N=1$ and bootstrapped with the two-gap approach. Figure \ref{N_1_result} is the result of $N=1$ bootstrap with two different values of $\Delta_{gap} = 4.5$ and $5.0$. The result indicates that the two sharp tip structure that we have observed at large $N$ persist to this smallest value.   The upper tip is still locked to the infrared Gaussian fixed point as for large $N$. The lower tip is still sharp, so we suspect that this lower tip represents an ultraviolet non-trivial fixed point of $N=1$ theory.

It is remarkable that $\Delta_{\sigma}$ values for the ultraviolet nontrivial fixed point do not vary widely for different theories. For all values of $N$ down to $O(1)$ at a fixed $\Delta_{\rm gap}$, $\Delta_{\sigma}$ hovers around the values $2.0 - 2.5$.
% what about N=0?

%---------------------------------------------------------
\subsubsection{Infrared Gaussian Fixed Point}
We also examined fine structure of the infrared Gaussian fixed point. Figure \ref{N_1_result} is the result for $N=1$ with $\Delta_{\rm gap} = 7.0$. Within the numerical precision, we find that the free field theory lies off the  upper tip location in the forbidden region. This indicates that such theory flows between two interacting fixed points of which the infrared one displays lower scaling dimension spectrum quite close to the free field theory.

To scrutinize our interpretation, we took an extreme limit $\Delta_{\rm{gap}}= 500.0$. The result is displayed in Figure \ref{N_1_result-2}. We see prominently that the tip location has considerably receded from the free theory location.

This pattern is intuitively clear. Nearly free field theory ought to contain a tower of scalar operators whose scaling dimensions take values close to integer multiple of $\Delta_{\boldsymbol \phi} = 1.5$. So, a theory which has no scalar operator between mid-gap $\Delta_\sigma \simeq 2 \Delta_{\boldsymbol \phi}$ and hierarchically large band-gap $
\Delta_{\rm gap} \gg \Delta_{\rm sigma}$ cannot possibly be a free or weakly interacting field theory. The extreme limit $\Delta_{\rm{gap}} \rightarrow \infty$ corresponds to a theory which contains only two states in the scalar operator spectrum. If such theory were to exist, the theory cannot be realized by a free or weakly interacting theory. Note, however, that this theory may (and in general does) have broad towers of higher-spin operator spectrum.

%%%%%%%%%%%%%%%%%%%%%%%%%%%%%%%%%%%%%%%%%%%%%%%%%%%%%%%%%%
\section{Conclusions and Discussions}

In this paper, we studied nontrivial conformal field theories in five dimensions in conformal bootstrap approach. Within the class of theories possessing $O(N)$ symmetry, we found convincing evidence for the existence of ultraviolet nontrivial fixed point.

For the purpose of fixed point identifications, we solved the conformal bootstrap conditions of the scalar sector.
For solving these conditions and, in particular, for identifying conformal field theories at the kink of allowed region boundary, we proposed what we termed two-gap approach. In this approach, above the identity operator, the theory features one scalar operator at the mid-gap $\Delta_{\rm min}$ and all other scalar operators start above the the band-gap $\Delta_{\rm gap}$. Thus, the conformal field theories we identified are specified by two parameters, $N$ and $\Delta_{\rm gap}$.

In the space of $(\Delta_{\boldsymbol \phi}, \Delta_{\rm min})$ of the scalar operators, we found universally that the boundary of bootstrap solutions features (at least) two tips, of which one is the value of free scalar field theory. At large $N$, the other tip (or a point close to it) asymptotes to the perturbative fixed point predicted by $1/N$ expansion. We took this as a convincing evidence that we correctly identified our conformal bootstrap result with the ultraviolet nontrivial fixed point.

As for the two-gap approach, finer specification of the spectrum would invariably pin down the candidate conformal field theories further. Yet, we found that the two-gap approach is sufficient enough for the purpose of identifying the location of nontrivial fixed points.
We relegate to a separate paper \cite{anotherpaper} our further investigation of two-gap approach and of multi-gap approach, and results for the spectrum of higher $O(N)$ representations, higher-spin operators and central charges.

We believe our finding bears strong implications in diverse fronts. Firstly, the existence of nontrivial $O(N)$ invariant conformal field theory in five dimensions
suggests that there may exist a vast class of higher-dimensional quantum field theories without the help of supersymmetry (as was for nontrivial fixed points in five and six dimensions) that exhibits the asymptotic safety.
It would be very interesting to explore them, especially including fermions and gauge fields, and chart out  classification of nontrivial ultraviolet fixed points.
It would also be interesting to consider these theories at continuous spacetime dimensions in the spirit similar to the $\epsilon$-expansion and the corresponding conformal bootstrap program.

Secondly, the nontrivial conformal field theory (other than the Gaussian fixed point) suggest exciting possibility of higher-spin gravity theory in six dimensions. In this regard, generalization of our finding to $U(N)$ invariant conformal field theory coupled to five dimensional gauge theory with or without Chern-Simons term would be extremely interesting. One would like to know if there are distinct classes of higher-spin gravity theories   with or without parity and time-reversal symmetries in a way reminiscent to the four-dimensional counterpart in higher-spin gravity \cite{Vasiliev:1999ba} and to the three-dimensional counterpart in bosonic \cite{Aharony:2011jz} or fermionic \cite{Giombi:2011kc} Chern-Simons matter theories, on which there is by now convincing checks from computations on both sides \cite{Giombi:2009wh}, \cite{Giombi:2010vg}, \cite{Giombi:2011ya}.
We are currently investigating these directions and will report results in the near future.

%%%%%%%%%%%%%%%%%%%%%%%%%%%%%%%%%%%%%%%%%%%%%%%%%%%%%%%%%%
\section*{Acknowledgement}
We thank Ofer Aharony, Dongsu Bak, John Cardy, Euihun Joung, Jaewon Kim, Zohar Komargodski, Hyunsoo Min, Sylvain Ribault, Woohyun Rim and Slava Rychkov for useful correspondences and discussions. We benefitted from stimulating scientific environment of the APCTP Focus Program "Liouville, Integrability and Branes (10)" during this work. We also acknowledge the KISTI Supercomputing Service Center for allocating runtime for the semi-definite programming routine. This research was supported in part by the National Research Foundation of Korea(NRF) grant funded by the Korea government(MSIP) through Seoul National University with grant numbers 2005-0093843, 2010-220-C00003 and 2012K2A1A9055280.

%%%%%%%%%%%%%%%%%%%%%%%%%%%%%%%%%%%%%%%%%%%%%%%%%%%%%%%%%%
\appendix

\section*{Appendix: Moduli Space of Conformal $n$-Points}
Table \ref{DOF} describes possible dynamical variables of $n$-point correlation function with respect to diverse spacetime dimensions. In this section, we describe details of that table. Let consider three-dimensional case. There is $SO(4,1)$ conformal symmetry with 10 generators. Correlation function ought be invariant under infinitesimal transformation generated by these generators. At the level of two point insertion, we can fix their location utilizing translation($P^\mu$) and special conformal($K^\mu$) generators. For convenience, we fix them on origin$(\vec{x_1} = O)$ and infinity$(\vec{x_4} =\infty)$. Therefore, 2-point correlation do not have dependence on conformal variables.

Suppose we add one more point inserted at $\vec{x_3}(r, \theta, \phi)$. Using dilatation generator, we can fix radial coordinate as $r=1$. Remaining two angle degrees could be fixed with part of rotation generator. Fixing these two angles $\theta,\phi$ means  specifying one axis and there is remained one rotation symmetry around this axis. Like 2-point, there is no possible conformal degree, 3-point function do not have dependence on conformal variables.

If we add 4-th point at $\vec{x_2}$, then remained one rotation symmetry used to fix one angle parameter of $\vec{x_2}$ and we have two kinematic variables correspond to unfixed degree of freedom. Therefore, 4-point correlation function should contain two kinematic variables, it would be realized via conformal cross ratios $u$ and $v$.

For the case of higher point insertion(larger than 4), no generators remained. Therefore number of kinematic variables just given by $3n-10$ when $n$ is larger than 4.\\

Let move on four-dimensional case and observe how many variables available for various $n$. Again, two points insertion at $\vec{x_1}$ and $\vec{x_4}$ could be fixed at origin and infinity with translation and special conformal symmetry. Likewise three dimension, there is no dependence on kinematic variables for 2-point correlation function.

To count moduli space of 3-point and 4-point correlation function, let parametrize $\vec{x_2}$ and $\vec{x_3}$,
\begin{align}
\vec{x_2} &= (r \ \mbox{cos}\theta_1, r \ \mbox{sin}\theta_1 \mbox{cos}\theta_2, r \ \mbox{sin}\theta_1 \mbox{sin}\theta_2 \mbox{cos}\theta_3, r \ \mbox{sin}\theta_1 \mbox{sin}\theta_2 \mbox{sin}\theta_3) \nonumber \\
\vec{x_3} &= (r' \ \mbox{cos}\theta_1', r' \ \mbox{sin}\theta_1' \mbox{cos}\theta_2', r' \ \mbox{sin}\theta_1' \mbox{sin}\theta_2' \mbox{cos}\theta_3', r' \ \mbox{sin}\theta_1' \mbox{sin}\theta_2' \mbox{sin}\theta_3')
\end{align}
$r, \theta_i$ are spherical coordinates. For $\vec{x_2}$, dilatation generator used to fix $r=1$ and three generator of rotation are used to fix $\theta_1,\theta_2,\theta_3$. We designate $x^1$ direction as the axis pass through origin and $\vec{x_2}$. That is, three rotation generators we used are $M_{12},M_{13},M_{14}$. All coordinates of $\vec{x_1},\vec{x_2},\vec{x_4}$ are fixed by conformal generator, again 3-point function do not depends on conformal variables.

Still, three rotation generators $M_{23},M_{24},M_{34}$ are remained. For fourth point insertion at $\vec{x_3}$, we nominate $x^2$ direction as the axis pass through origin and $\vec{x_3}$, without loss of generality. Specifying this axis, two generators $M_{23},M_{24}$ can be used to fix two radial coordinates $\theta_1',\theta_2'$ of $\vec{x_3}$. However, rotation around $x^3-x^4$ plane(generated by $M_{34}$) is cannot be used to break symmetry. Therefore, two degrees with respect to $r', \theta_3'$ remains and this explains why 2 cross ratios should be appear on 4-point correlation function in 4-dimension. More higher point, all conformal generators utilized to break symmetry, therefore number of variables given by $4n-15$. \\

Higher dimension generalization is straightforward, which result displayed in Table \ref{DOF}. As far as spacetime dimension is larger than 2, there should be 2 variables on 4-point correlation function. Therefore, we can apply 1-parameter or 2-parameter bootstrap process regardless of spacetime dimension.


\newpage

\begin{thebibliography}{999}

\bibitem{Polyakov:1974gs}
  A.~M.~Polyakov,
  ``Nonhamiltonian approach to conformal quantum field theory,''
  Zh.\ Eksp.\ Teor.\ Fiz.\  {\bf 66} (1974) 23.
  %%CITATION = ZETFA,66,23;%%

\bibitem{Belavin:1984vu}
  A.~A.~Belavin, A.~M.~Polyakov and A.~B.~Zamolodchikov,
  ``Infinite Conformal Symmetry in Two-Dimensional Quantum Field Theory,''
  Nucl.\ Phys.\ B {\bf 241} (1984) 333.
  %%CITATION = NUPHA,B241,333;%%

\bibitem{Rattazzi:2008pe}
  R.~Rattazzi, V.~S.~Rychkov, E.~Tonni and A.~Vichi,
  ``Bounding scalar operator dimensions in 4D CFT,''
  JHEP {\bf 0812} (2008) 031
  [arXiv:0807.0004 [hep-th]].
  %%CITATION = ARXIV:0807.0004;%%

\bibitem{Poland:2010wg}
  D.~Poland and D.~Simmons-Duffin,
  ``Bounds on 4D Conformal and Superconformal Field Theories,''
  JHEP {\bf 1105} (2011) 017
  [arXiv:1009.2087 [hep-th]].
  %%CITATION = ARXIV:1009.2087;%%

\bibitem{Rychkov:2009ij}
  V.~S.~Rychkov and A.~Vichi,
  ``Universal Constraints on Conformal Operator Dimensions,''
  Phys.\ Rev.\ D {\bf 80} (2009) 045006
  [arXiv:0905.2211 [hep-th]].
  %%CITATION = ARXIV:0905.2211;%%

\bibitem{Caracciolo:2009bx}
  F.~Caracciolo and V.~S.~Rychkov,
  ``Rigorous Limits on the Interaction Strength in Quantum Field Theory,''
  Phys.\ Rev.\ D {\bf 81} (2010) 085037
  [arXiv:0912.2726 [hep-th]].
  %%CITATION = ARXIV:0912.2726;%%

\bibitem{Rattazzi:2010gj}
  R.~Rattazzi, S.~Rychkov and A.~Vichi,
  ``Central Charge Bounds in 4D Conformal Field Theory,''
  Phys.\ Rev.\ D {\bf 83} (2011) 046011
  [arXiv:1009.2725 [hep-th]].
  %%CITATION = ARXIV:1009.2725;%%

\bibitem{ElShowk:2012ht}
  S.~El-Showk, M.~F.~Paulos, D.~Poland, S.~Rychkov, D.~Simmons-Duffin and A.~Vichi,
  ``Solving the 3D Ising Model with the Conformal Bootstrap,''
  Phys.\ Rev.\ D {\bf 86} (2012) 025022
  [arXiv:1203.6064 [hep-th]].
  %%CITATION = ARXIV:1203.6064;%%

\bibitem{El-Showk:2014dwa}
  S.~El-Showk, M.~F.~Paulos, D.~Poland, S.~Rychkov, D.~Simmons-Duffin and A.~Vichi,
  ``Solving the 3d Ising Model with the Conformal Bootstrap II. c-Minimization and Precise Critical Exponents,''
  J.\ Stat.\ Phys.\ xx {\bf } (2014) xx
  [arXiv:1403.4545 [hep-th]].
  %%CITATION = ARXIV:1403.4545;%%

\bibitem{Joung:2012a}
  X.~Bekaert, E.~Joung and J.~Mourad,
  ``Comments on higher-spin holography'',
  arXiv:1202.0543 [hep-th].

\bibitem{Sezgin:2002rt}
  E.~Sezgin and P.~Sundell,
  ``Massless higher spins and holography,''
  Nucl.\ Phys.\ B {\bf 644} (2002) 303
   [Erratum-ibid.\ B {\bf 660} (2003) 403]
  [hep-th/0205131].
  %%CITATION = HEP-TH/0205131;%%

\bibitem{Klebanov:2002a}
  I.~Klebanov and A.~M.~Polyakov,
  ``AdS dual of the critical $O(N)$ vector model'',
  arXiv:0210114 [hep-th].

\bibitem{Henneaux:2010xg}
  M.~Henneaux and S.~J.~Rey,
  ``Nonlinear $W_{infinity}$ as Asymptotic Symmetry of Three-Dimensional Higher Spin Anti-de Sitter Gravity,''
  JHEP {\bf 1012} (2010) 007
  [arXiv:1008.4579 [hep-th]].
  %%CITATION = ARXIV:1008.4579;%%

\bibitem{Klebanov:2014a}
  L.~Fei, S.~Giombi and I.~Klebanov,
  ``Critical $O(N)$ Models in $6-\epsilon$ Dimensions'',
  arXiv:1404.1094 [hep-th].

\bibitem{Klebanov:2014b}
  L.~Fei, S.~Giombi, I.~Klebanov and G.~Tarnopolsky,
  ``Three Loop Analysis of the Critical $O(N)$ Models in $6-\epsilon$ Dimensions'',
  arXiv:1411.1099 [hep-th].

\bibitem{Nakayama:2014a}
  Y.~Nakayama and T.~Ohtsuki,
  ``Five dimensional $O(N)$-symetruc CFTs from conformal bootstrap'',
  arXiv:1404.5201 [hep-th].

\bibitem{Simmons-Duffin:2013a}
  F.~Kos, D.~Poland and D.~Simmons-Duffin,
  ``Bootstrapping the $O(N)$ Models'',
  arXiv:1307.6856 [hep-th].

\bibitem{Dolan:2001a}
  F.~A.~Dolan and H.~Osborn,
  ``Conformal Four Point Functions ans the Operator Product Expansion'',
  arXiv:0011040 [hep-th].

\bibitem{Dolan:2011a}
  F.~A.~Dolan and H.~Osborn,
  ``Conformal Partial Waves : Futher Mathematical Results'',
  arXiv:1108.6194 [hep-th].

\bibitem{Ferrara:1971a}
  S.~Ferrara, R.~Gatto and A.~F.~Grillo,
  ``Conformal Invariance on the Light cone and Canonincal Dimensions'',
  Nucl. Phys. B34(1971), p.349.

\bibitem{Mack:1977a}
  G.~Mack,
  ``Convergence of Operator Product Expansions on the Vacuum in Conformal Invariant Quantum Field Theory'',
  Commun. math. Phys53, 155-184(1977).

\bibitem{Rychkov:2012a}
  S.~El-Showk, M.~F.~Paulos, D.~Poland, S.~Rychkov, D.~Simmons-Duffin and A.~Vichi,
  ``Solving the 3D Ising Model with the Conformal Bootstrap'',
  arXiv:1203.6064 [hep-th].

\bibitem{Polyakov:1974a}
  A.~M.~Polyakov,
  ``Non-Hamiltonian approach to conformal quantum field theory'',
  Zh. Eksp. Teor. Fiz. 66, 23-42(1974).

\bibitem{Rychkov:2008a}
  R.~Rattazzi, S.~Rychkov, E.~Tonni and A.~Vichi,
  ``Bounding scalar operator dimensions in 4D CFT'',
  arXiv:0807.0004 [hep-th].

\bibitem{Simmons-Duffin:2011a}
  D.~Poland, D.~Simmons-Duffin and A.~Vichi,
  ``Carving Out the Space of 4D CFTs'',
  arXiv:1109.5176 [hep-th].

\bibitem{Rychkov:2013a}
  M.~Hogervorst and S.~Rychkov,
  ``Radial Coordinates for Conformal Blocks'',
  arXiv:1303.1111 [hep-th].

\bibitem{Gliozzi:2014a}
  F.~Gliozzi and A.~Rago,
  ``Critical exponents of the 3d Ising and related models from Conformal Bootstrap'',
  arXiv:1403.6003 [hep-th].



\bibitem{Vichi:2011a}
  A.~Vichi,
  ``Improved bounds for CFT's with global symmetries'',
  arXiv:1106.4037 [hep-th].

\bibitem{Kotikov:1996a}
  D.~J.~Broadhurst and A~.V.~Kotikov,
  ``Compact anayltical form for non-zeta terms in critical exponents at order $\frac{1}{N^3}$'',
  arXiv:1106.4037 [hep-th].


\bibitem{anotherpaper}
J.~Bae, J.~W. Kim and S.J.~Rey, to appear (2015).


\bibitem{Vasiliev:1999ba}
  M.~A.~Vasiliev,
  ``Higher spin gauge theories: Star product and AdS space,''
  In *Shifman, M.A. (ed.): The many faces of the superworld* 533-610
  [hep-th/9910096].
  %%CITATION = HEP-TH/9910096;%%


\bibitem{Aharony:2011jz}
  O.~Aharony, G.~Gur-Ari and R.~Yacoby,
  ``d=3 Bosonic Vector Models Coupled to Chern-Simons Gauge Theories,''
  JHEP {\bf 1203} (2012) 037
  [arXiv:1110.4382 [hep-th]].


\bibitem{Giombi:2011kc}
  S.~Giombi, S.~Minwalla, S.~Prakash, S.~P.~Trivedi, S.~R.~Wadia and X.~Yin,
  ``Chern-Simons Theory with Vector Fermion Matter,''
  Eur.\ Phys.\ J.\ C {\bf 72} (2012) 2112
  [arXiv:1110.4386 [hep-th]].


\bibitem{Giombi:2009wh}
  S.~Giombi and X.~Yin,
  ``Higher Spin Gauge Theory and Holography: The Three-Point Functions,''
  JHEP {\bf 1009} (2010) 115
  [arXiv:0912.3462 [hep-th]].
  %%CITATION = ARXIV:0912.3462;%%

\bibitem{Giombi:2010vg}
  S.~Giombi and X.~Yin,
  ``Higher Spins in AdS and Twistorial Holography,''
  JHEP {\bf 1104} (2011) 086
  [arXiv:1004.3736 [hep-th]].
  %%CITATION = ARXIV:1004.3736;%%

\bibitem{Giombi:2011ya}
  S.~Giombi and X.~Yin,
  ``On Higher Spin Gauge Theory and the Critical O(N) Model,''
  Phys.\ Rev.\ D {\bf 85} (2012) 086005
  [arXiv:1105.4011 [hep-th]].
  %%CITATION = ARXIV:1105.4011;%%

\end{thebibliography}
\end{document}